\documentclass[structabstract]{aa}  
\usepackage{dcolumn}
\usepackage[T1,T5]{fontenc}
\usepackage{graphicx}
\usepackage{txfonts}
\usepackage[normalem]{ulem}
\usepackage{natbib}
\bibpunct{(}{)}{;}{a}{}{,}

\newcommand{\cmc}{\mbox{$\mbox{cm}^{-3}$}}
\newcommand{\kms}{\mbox{km$\,$s$^{-1}$}}
\newcommand{\cmg}{\mbox{$\mbox{cm}^{2} \, \mbox{g}^{-1}$}} 
\newcommand{\lsun}{\mbox{$L_\odot$}}
\newcommand{\msun}{\mbox{$M_\odot$}}
\newcommand{\mjb}{\mbox{$\mbox{mJy} \, \mbox{beam}^{-1}$}}

\newcommand{\K}{\mbox{K}}

\newcommand{\micron}{\mbox{$\mu$m}}

\newcommand{\vlsr}{\mbox{$V_{\mbox{\tiny LSR}}$}}
\newcommand{\tmb}{\mbox{$T_{\mbox{\tiny MB}}$}}

\newcommand{\tdust}{\mbox{$T_{\mbox{\tiny dust}}$}}
\newcommand{\kmm}{\mbox{$\kappa_{\mbox{\tiny 870~$\mu$m}}$}}

\begin{document}
   \title{W43: the closest molecular complex of the Galactic Bar?\thanks{Figures A1-4 in appendix A are only available
in electronic form at http://www.aanda.org}}
\author{Q. Nguy\~\ecircumflex n L\uhorn \ohorn ng\inst{1}
          \and F. Motte\inst{1}
          \and F. Schuller\inst{2}
          \and N. Schneider\inst{1}    
          \and S. Bontemps\inst{3}
          \and P. Schilke\inst{4}
          \and K. M. Menten\inst{2}
          \and F. Heitsch\inst{5}
          \and F. Wyrowski\inst{2}
          \and P. Carlhoff\inst{4}
          \and L. Bronfman\inst{6}
          \and T. Henning\inst{7}         
          }
          
  \institute{Laboratoire AIM Paris-Saclay, CEA/IRFU - CNRS/INSU - Universit\'e Paris Diderot, Service d'Astrophysique, B\^at. 709, CEA-Saclay, F-91191 Gif-sur-Yvette Cedex, France, 
              \email{ quang.nguyen-luong@cea.fr}
      \date{Received 06 Dec 2010; accepted 14 Feb 2011}
          \and Max-Planck-Institut f\"ur Radioastronomie, Auf dem H\"ugel 69, D-53121 Bonn, Germany
         \and OASU/LAB-UMR~5804, CNRS/INSU - Universit\'e Bordeaux 1, 2 rue de l'Observatoire, BP 89, F-33270 Floirac, France
        \and I. Physikalisches Institut, Universit\"at zu K\"oln, Z\"ulpicher Str. 77, 50937 K\"oln, Germany
        \and Department of Physics and Astronomy, University of North Carolina Chapel Hill, Phillips Hall, Chapel Hill, NC 27599-3255, USA
        \and Departamento de Astronomia, Universidad de Chile, Casilla 36-D, Santiago, Chile
        \and Max Planck Institute for Astronomy, Koenigstuhl 17, D-69117 Heidelberg, Germany        
        }
   \abstract
   {} 
   {In the framework of multi-wavelength Galactic surveys of star formation which are presently underway, complexes of molecular clouds that stretch over up to hundreds of parsecs are of particular interest. This is because a large population of stars are forming inside them, thus at the same distance from the Sun and under similar physical conditions. In the present paper, we focus on the range of the Galactic plane between $\approx 29.5$ and 31.5 degrees of longitude which is especially rich in terms of molecular clouds and star formation activity. It is located within what is sometimes called the Molecular Ring and it contains the Galactic mini-starburst region W43 as well as the prominent hot core G29.96-0.02 with its associated compact H~{\scriptsize II} region. }
   {We used a large database extracted from Galaxy-wide surveys of H~{\scriptsize I}, $^{13}$CO~1-0, $8~\micron$ and $870~\micron$ continuum to trace diffuse atomic gas, low- to medium-density molecular gas, high-density molecular gas, and star formation activity which we complemented by dedicated $^{12}$CO~2--1, 3--2 observations of the region.}
   {From the detailed 3D (space-space-velocity) analysis of the molecular and atomic cloud tracers through the region and despite its wide velocity range (\emph{FWHM}$\sim22.3~ \kms$ around $\vlsr\sim 95.9~\kms$), we identified W43 as a large (equivalent diameter $\sim 140$~pc) and coherent complex of molecular clouds which is surrounded by an atomic gas envelope (equivalent diameter $\sim 290$~pc). We measured the total mass of this newly-identified molecular complex ($M_{\mbox{\tiny total}}\sim 7.1 \times 10^6~\msun$), the mass contained in dense $870~\micron$  clumps ($<5$~pc dense cloud structures, $M_{\mbox{\tiny clumps}}\sim 8.4 \times 10^5~\msun$) and conclude that W43 is particularly massive and concentrated. The distance we assume for the W43 complex is  6~kpc from the Sun, which may place it at the meeting point of the Scutum-Centaurus (or Scutum-Crux) Galactic arm and the Bar, a dynamically complex region where high-velocity streams could easily collide. The star formation rate of W43 is suggested not to be steady but it is increasing from $\, \sim 0.01~\msun\, \mbox{yr}^{-1}$ (measured from its $8\,\mu$m luminosity) to $\sim 0.1~\msun\, \mbox{yr}^{-1}$ (measured from its molecular content). From the global properties of W43, we claim it is an extreme molecular complex in the Milky Way and it could even be forming starburst clusters in the near future.}
  {W43 is the perfect testbed to investigate (1) the star formation process happening through bursts as well as (2) the formation of such an extreme complex in the framework of converging flows scenarios.}
\titlerunning{}
\authorrunning{Nguy\~\ecircumflex n L\uhorn \ohorn ng et al.}
\maketitle
%

\begin{table*}[htbp]
\caption{Observational parameters of the tracers used in the present paper.}
\label{table:obs} 
\centering 
\begin{tabular}{l l l l l l l l} 
\hline
\hline
Telescope 					&  Tracer				 & Frequency	&    $HPBW$  & Velocity Range & $\Delta$ v$_{\mbox{\tiny res}}$ & 1$\sigma$~Rms\\   
$\diagup$Survey    				&					 &     ~~~[GHz] 	&    [arcsec]    & ~~~~~~[$\kms$] & [$\kms$] &    \\   
\hline
VLA$\diagup$VGPS				& H \scriptsize{I}	   	 & ~~~~1.420    &~~60\arcsec	& -120 to 170	& 1.56	& 1.80 K$\,\kms$  \\
CfA$\diagup$CfA		            	& $^{12}$CO~1--0     		 & 115.271 	& 450\arcsec	& ~-0.5 to 271 	& 0.65 	& 0.22 K$\,\kms$   \\
KOSMA$\diagup$present paper	& $^{12}$CO~2--1     		 & 230.537	& 130\arcsec	& ~~~~~0 to 200		& $0.20$	&  0.11 K$\,\kms$\\
KOSMA$\diagup$present paper	& $^{12}$CO~3--2     		 & 345.796	& ~~80\arcsec	&  ~~~~~0 to 200	& $0.30$	&  0.52 K$\,\kms$ \\
FCRAO$\diagup$GRS			& $^{13}$CO~1--0   	  	 & 110.201	& ~~46\arcsec	&~~~~-5 to 135		& $0.21$	& 0.13 K$\,\kms$ \\ 
\hline
APEX$\diagup$ATLASGAL 		& 870 $\mu$m continuum & ... 			& ~~19\arcsec  	& ...		         & ...	         & $60~\mjb$\\   
SPITZER$\diagup$GLIMPSE 		& 8 $\mu$m continuum	 & ...			&   ~~~6\arcsec  	& ...		         & ... 	         & $0.08~\mjb$\\  
\hline
\end{tabular}
\end{table*}

\section{Introduction}
Molecular clouds forming low-mass stars in our solar neigborhood have been well known and are mostly located in the Gould Belt, a giant coherent cloud structure inclined to the Galactic plane (e.g. \citealt{comeron94}). High-mass stars should form in complexes of Giant Molecular Clouds (GMCs) also called Giant Molecular Associations (GMAs), i.e. groups of molecular clouds with a total mass maybe up to $10^{7}~\msun$ \citep{rand90,rand93b,kuno95,koda09}. While a few molecular complexes have been recognized to be forming high-mass stars in the outer Galaxy (e.g. Cygnus~X and, in a more modest form, NGC~7538), finding coherent associations of high-mass star-forming regions on a Galactic scale has been challenging. The main reason for this is that near the Galactic equator, where these massive complexes are expected to reside, the identification of molecular complexes/GMAs is confused by other line-of-sight clouds. Galaxy-wide low-resolution CO surveys have provided lists of molecular cloud groups which are often interpreted as consisting of a blend of clouds spreading along the line of sight  (e.g. \citealt{dame01}). We hereafter call W43 the emission threading the region l=(29\degr -- 32\degr), b=(-1\degr -- +1\degr) of the Galactic plane. W43 has been identified to contain two of the largest cloud groups of the first Galactic Quadrant with a total mass of $\sim 5 \times 10^6~\msun$~ (close to l=$31^\circ$ and line-of-sight velocity, $\vlsr = 95~\kms$, and close to l=$29^\circ$ and $\vlsr = 80~\kms$; \citealt{dame86}). 
\cite{solomon87} used higher-resolution CO surveys to split the ensemble of W43 clouds into 14 clouds and locate the 2 most massive ones: SRBY~162  (which we call \object{W43-Main}) and SRBY~171 (which is called \object{W43-South}) with virial masses of several times $10^6~\msun$ (see also \citealt{liszt95}; \citealt{mooney95}). This is confirmed by \cite{rathborne09} who used $^{13}$CO emission as a higher-density tracer observed in the higher resolutions Galactic Ring Survey (GRS) to identify more than 20 molecular clouds in the W43 region, with systemic velocities ranging from $12~\kms$ to $110~\kms$.
The core of W43-Main harbors a well-known giant H~{\scriptsize II}  region powered by a particularly luminous cluster of Wolf-Rayet (WR) and OB stars ($\sim 3.5\times 10^{6}~ \lsun$, see \citealt{blum99} and references therein). \cite{motte03} established it as a dense region equivalent to a mini-starburst region since it is undergoing a remarkably efficient episode of high-mass star formation (the discovery of $\sim 15$ high-mass protoclusters leading to star formation efficiency (\emph{SFE}) of ~$\sim 25\%/10 ^6$~yr). Recent far-infrared to submillimeter data from the \emph{Herschel} Space Observatory revealed a complex structure of chimneys and filaments, and confirmed its efficiency in forming massive stars \citep{bally10a}.  W43-South corresponds to a  less extreme cloud which however harbors the well-known compact H~{\scriptsize II} (CH~{\scriptsize II}) region\footnote{G29.96-0.02 is normally termed an ultracompact H{~\scriptsize II} region. However, we note that its size and radio flux density resemble that of the Orion Nebula, the archetypical CH{~\scriptsize II} region, when the latter is put at a distance of 6~kpc. G29.96-0.02 excited by at least one O star (\citealt{cesaroni98}; \citealt{pratap99}; \citealt{beuther07})}. The environment of W43-Main and W43-South requires more investigation to check if these GMCs could be related and form an individual complex/GMA.

The APEX Telescope Large Area Survey of the Galaxy (ATLASGAL, \citealt{schuller09}) has made an unbiased census of young stellar objects deeply embedded in the GMCs. The survey of the inner Galactic plane including the W43 region displays a rich collection of compact sources within W43, which correspond to the high-density backbones of molecular clouds. 

Before making a meaningful assessment of the global characteristics of the star-forming regions, a strong effort is needed to identify GMCs and groups of GMCs which we could call a molecular complex or GMA. In this paper, we define the W43 molecular complex and then determine its global characteristics. Employing the large database described in Sect.~\ref{obs}, we show that W43 is a coherent molecular and star-forming complex surrounded by atomic gas (see Sect.~\ref{results}). In Sect.~\ref{discussion}, we locate W43 at the connecting point between the Scutum-Centaurus arm and the Galactic Bar and quantify its mass and star formation activity. Section~5 concludes that W43 is indeed one extreme molecular cloud and star-forming complex of our Galaxy, located at only $\sim$~6~kpc from the Sun.

\section{Observations and database}
\label{obs}
We used data from different Galactic plane surveys and performed additional observations of the W43 region. These data sets are summarized in Table~\ref{table:obs}.
\subsection{H~{\scriptsize I} database}

The atomic gas data of W43 are taken from the Very Large Array (VLA) Galactic Plane Survey (VGPS\footnote{
	The full spectral cubes of the VGPS database are available as fits files at \emph{http://www.ras.ucalgary.ca/VGPS/}. The VLA and GBT are facilities of the National Radio Astronomy Observatory.},
\citealt{stil06}). It is a survey of the 21 cm continuum and line emission from neutral atomic hydrogen, H~{\scriptsize I}, performed through the Galactic plane ($18\degr < l < 67\degr$, $|b| < 1.3\degr$) by the VLA and combined with short-spacing information obtained with the Green Bank Telescope (GBT). This data set has a spatial resolution of $1\arcmin\times1\arcmin$ and a velocity resolution of $1.56~\kms$ spanning the range -120 to $170~\kms$.

\begin{figure*}[ht]
\vskip -0.5cm
\centering
\includegraphics[angle=-90,width=17.cm]{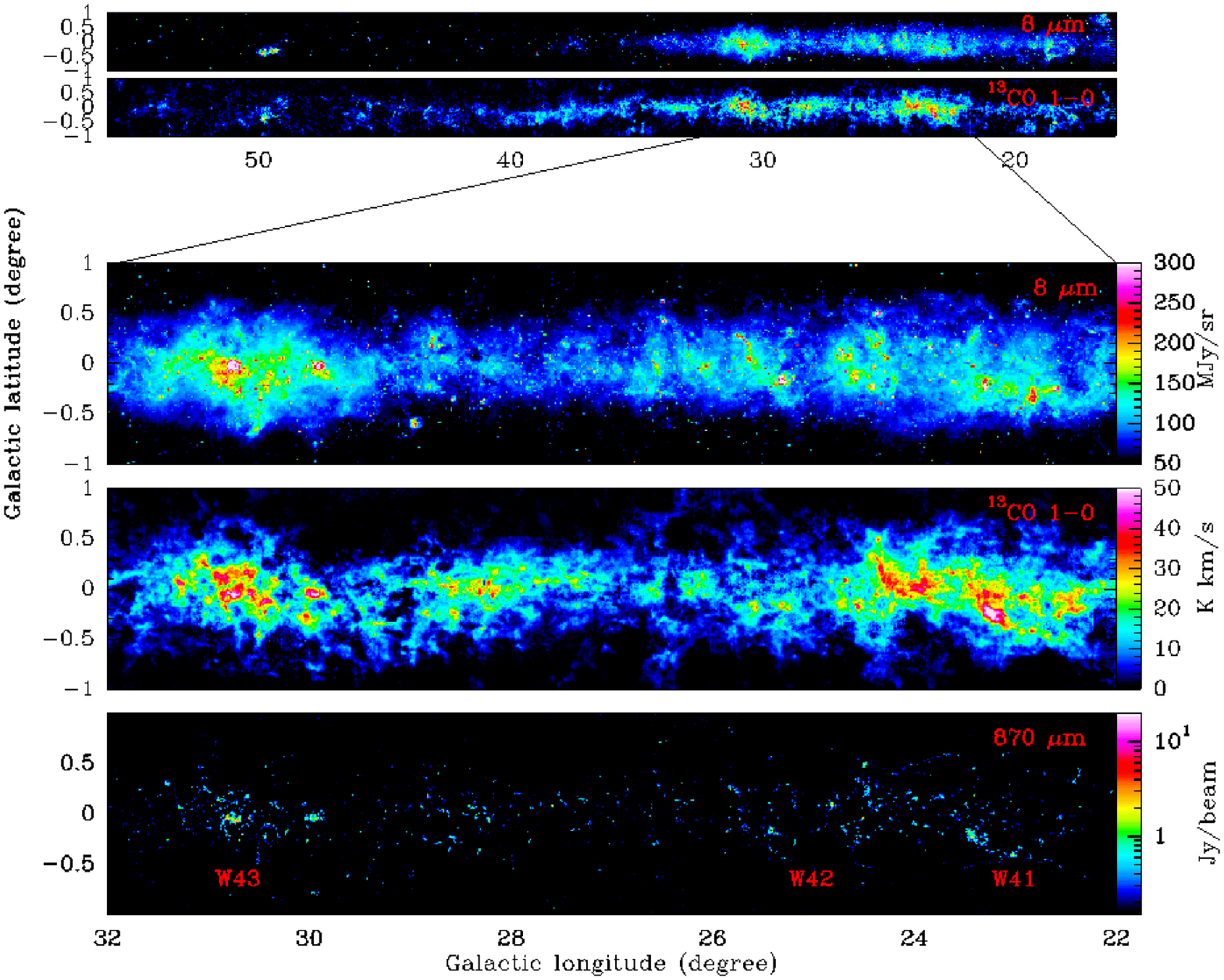} 
\vskip -0.3cm
\caption{Galactic plane from $l=18\degr$ to $l=56\degr$ shown: {\bf a)} at $8\,\mu$m \citep{benjamin03}; {\bf b)} in the $^{13}$CO~1--0  line, integrated from $-5$ to $135~\kms$ \citep{jackson06}. Zoom on $l=22\degr-32\degr$: {\bf c)} at $8\,\mu$m; {\bf d)} in $^{13}$CO~1--0 ; and {\bf e)} at $870\,\mu$m \citep{schuller09}. }
\label{fig:GP}
\vskip -0.5cm
\end{figure*}

\subsection{$^{12}$CO observations and $^{13}$CO database}
To trace the low-density molecular gas, we used the  $^{12}$CO~1--0  and $^{13}$CO~1--0  data from previous Galactic plane surveys and, additionally,  observed the $^{12}$CO~2--1  and $^{12}$CO~3--2  lines  with the 3 m diameter K\"olner Observatorium f\"ur Submm-Astronomie (KOSMA\footnote{
	The KOSMA 3 m submillimeter telescope was installed on Gornergrat-S\"ud, Switzerland, operated by the University of Cologne in collaboration with the University of Bonn. It is now rebuilt close to Lhasa/Tibet.}) from 2009 January to February.

The $^{12}$CO~1--0 data set is taken from the Galactic plane survey made with the CfA 1.2 m telescope\footnote{The CfA telescope is  the 1.2 m Millimeter-Wave Telescope at the Center for Astrophysics, Harvard University.}  \citep{dame01}. These data have a  spatial resolution of $450\arcsec$ and a velocity resolution of $0.65~\kms$ spanning the velocity range -0.5 to $271~\kms$.

We also used the $^{13}$CO~1--0 data of the Boston University- Five College Radio Astronomy Observatory (BU-FCRAO) Galactic Ring Survey (GRS)\footnote{
The full spectral fits cubes are available to download at  \emph{http://www.bu.edu/galacticring}.}. 
This survey was performed with the FCRAO 14 m telescope and covers the Galactic plane in the range $18\degr < l < 55.7\degr$ for  $|b| < 1\degr$ \citep{jackson06}. It has a spatial resolution of 46$\arcsec$ and a velocity resolution of $0.21~\kms$ spanning the velocity range -5 to $135~\kms$. The spectral cubes are converted into a main-beam brightness temperature using an efficiency of  0.48. Such a calibration can overestimate the brightness by up to 30\% since stray radiation is not yet corrected for (Chris Brunt priv. com.).

Additionally, observations of $^{12}$CO~2--1  and $^{12}$CO~3--2  lines were performed with a dual-channel SIS-receiver, built at KOSMA, operating at 210-270 and 325-365 GHz, that is mounted on the KOSMA telescope (\citealt{graf98}). The spectra were calibrated to a main-beam brightness temperature scale ($T_{\mbox{\tiny MB}}$) using the main beam efficiencies 0.72 (230~GHz) and 0.68 (345~GHz). Pointing was monitored simultaneously for both receiver channels using continuum cross scans of Jupiter and was found to be accurate to within $15\arcsec-30\arcsec$. The calibration accuracy is approximately 20\%. We used the on-the-fly observing mode at KOSMA (\citealt{kramer98b}) to map and resample areas of $\sim 1\degr^{2}$ (for $^{12}$CO~2--1 ) and $\sim 0.5\degr^{2}$ (for $^{12}$CO~3--2 ) on $60\arcsec\times60\arcsec$ grids. The angular resolution at 230 GHz and at 345 GHz are 130$\arcsec$ and 80$\arcsec$, respectively. Figures. A1-4 show the integrated and channel maps of $^{12}$CO~2--1  and $^{12}$CO~3--2  emissions obtained with KOSMA. Note that only the region around W43-Main was mapped in the $^{12}$CO~3--2  emission line.

\subsection{Dust continuum image at $870\,\mu$m}

 The APEX\footnote{
APEX, the Atacama Pathfinder Experiment, is a collaboration between the Max Planck Institut f\"ur Radioastronomie, the Onsala Space Observatory  and the European Southern Observatory. Images are available at  \emph{http://www.mpifr-bonn.mpg.de/div/atlasgal/}.} telescope large area survey of the Galaxy (ATLASGAL, \citealt{schuller09}) provides an image of the 870~$\mu$m dust continuum emission of W43 as part of its coverage of the inner Galactic plane ($-60\degr~<~l~<~60\degr$ for  $|b|~<~1-1.5\degr$) imaging. This survey was completed with the Large APEX Bolometer Camera Array (LABOCA) installed at the 12~m Atacama Pathfinder Experiment (APEX) telescope located on Llano Chajnantor in Chile.
 Its bandpass is centered at 345~GHz and has a bandwidth of 60~GHz. The \emph{FWHM} at this frequency is $\sim 19\arcsec$ and the mean rms noise level in the W43 region is $\sim60~\mjb$. The data reduction is described in \cite{schuller09} and has been presently improved. In the ATLASGAL images, most of the emission that appears uniform on scales larger than $6\arcmin$ is effectively filtered out. At 6~kpc and above the $5\sigma$ level, the ATLASGAL survey is tracing $> 80~\msun$ compact cloud fragments (see Eq. \ref{eq:mclump}) that have sizes $< 5$~pc, and thus volume density $> 2 \times 10^3~\cmc$.

\subsection{Mid-infrared database}

W43 was imaged by the \emph{Spitzer} satellite as part of the Legacy Science Program GLIMPSE (Galactic Legacy Infrared Mid-Plane Survey Extraordinaire infrared survey) which covers the Galactic plane ($-65\degr < l~<~65\degr$ for  $|b|~<~1\degr$, \citealt{benjamin03}). The emission in the  \emph{Spitzer}/IRAC band 4 at 8~$\mu$m is dominated by emission of polycyclic aromatic hydrocarbon (PAH) heated by UV photons and is thus related to the ``present" star formation activity (e.g. \citealt{peeters04}).

\section{Results and analysis}
\label{results}

\subsection{W43: a prominent region of the first Galactic Quadrant}
\label{Q1}
High-angular-resolution surveys of the Galactic plane have generally focused on the $-60^\circ<l<60^\circ$ longitude range (e.g. ATLASGAL and GLIMPSE) or even avoiding the Galactic center as in the GRS survey ($18^\circ<l<56^\circ$). Figure~\ref{fig:GP} presents the Galactic plane from $l=18\degr$ to $56\degr$ as seen in different surveys, tracing the column density of molecular clouds ($^{13}$CO), their star formation activity (8~$\mu$m), and the regions of high-density gas ($870\,\mu$m). In the intensity map in which the $^{13}$CO~1--0  emission is integrated over the complete velocity range it covers (from -5 to 135~$\kms$, see Fig.~\ref{fig:GP}), W43 indeed stands out as one of the largest and brightest groups of extended molecular complexes extending over $\sim 2$ degrees around Galactic longitude $l=30.5\degr$. Moreover, in the $870\,\mu$m and $8\,\mu$m surveys which trace dense protostellar material and activity of newly born stars, respectively, W43 is also one of the brightest regions of the first Galactic Quadrant (see Fig. \ref{fig:GP}). With its molecular gas brightness, concentration of cloud material in dense regions, and its high star formation activity, W43 is one of the most outstanding complexes in the first Quadrant. 

\subsection{The main velocity range of W43}
\label{velrange}
\begin{figure}[b]
\vskip -0.2cm
\centering
\includegraphics[angle=-90,width=8.7cm]{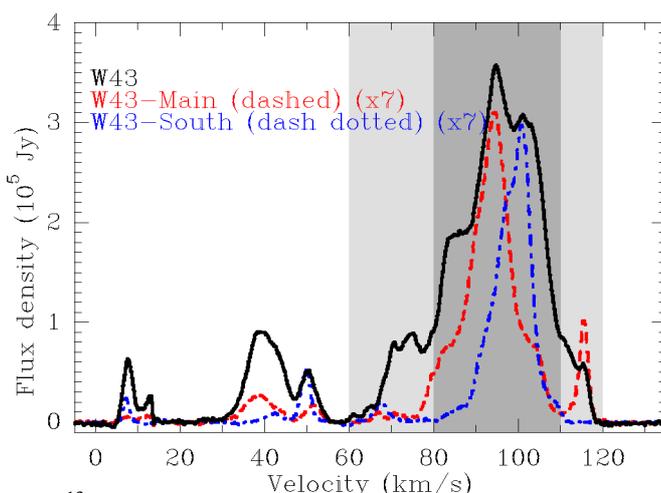} 
\vskip -0.3cm
\caption{$^{13}$CO~1--0  line spectra resulting from the sum of all spectra observed in the $1.8\degr\times 0.8\degr$ area covering W43 (black line), compared to those summed in the circular areas shown in Fig.~\ref{ATLASGAL} to be associated with W43-Main (red dashed line) and W43-South (blue dashed-dotted line). The main and complete velocity ranges of W43 (see text) are indicated with light- and dark-gray filling.}
\label{13cospec}
\vskip -0.5cm
\end{figure}
\begin{figure}[!h]
\vskip -0.5cm
\includegraphics[angle=0,height=22.1cm]{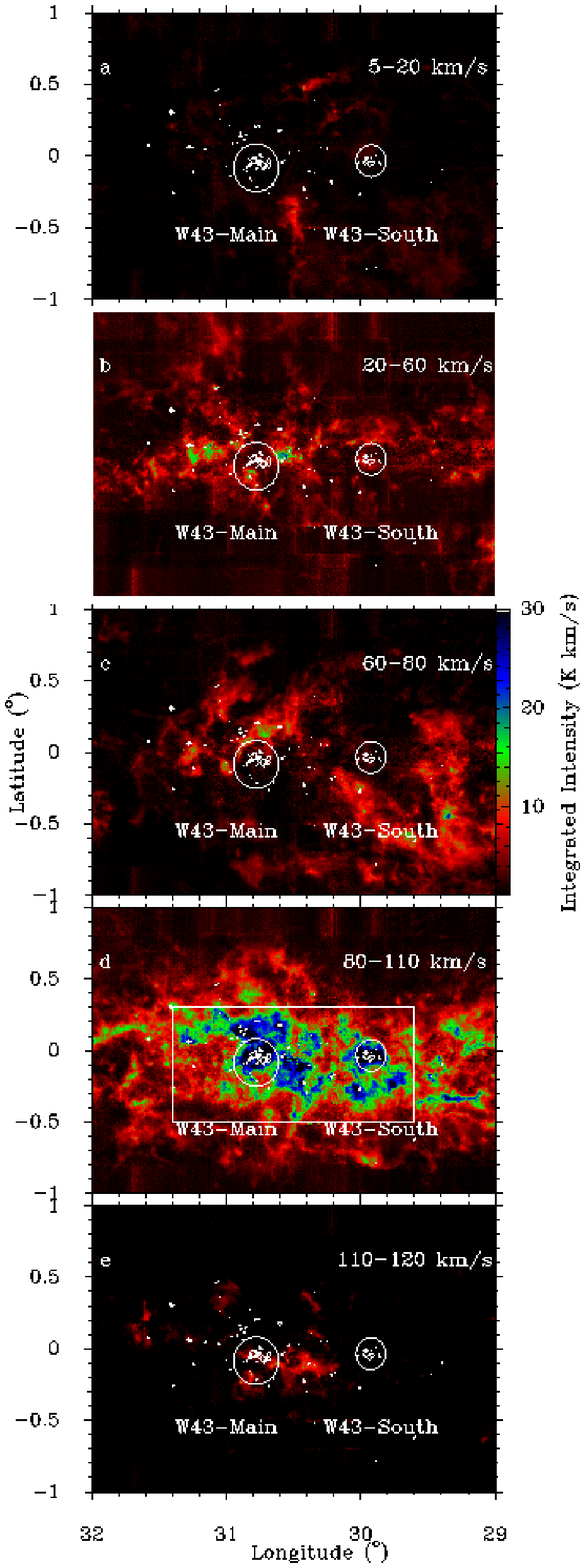} 
\caption{Integrated maps of the $^{13}$CO~1--0  line (in color scale) in the direction of the W43 molecular cloud complex whose densest parts are outlined by the 870~$\mu$m continuum emissions (in contours). The various velocity intervals used for the integration correspond to : {\bf a)-b)} clouds not associated with W43; {\bf c), e)} clouds emitting in the low- and high-velocity wings of the W43 line shown in Fig.~\ref{13cospec}; {\bf d)} the main velocity range of the W43 molecular complex. The white circles locate W43-Main and W43-South, their areas are used to integrate the spectra of Fig.~\ref{13cospec}.} 
\vskip -1.3cm
\label{f:13co_chanmaps}
\end{figure}

  
Though the image of the $^{13}$CO integrated intensity shows W43 as a prominent region in this portion of the Galactic plane, the velocity range from $-5$ to $135~\kms$ certainly covers clouds lying at various distances along the same line of sight. This becomes obvious when inspecting the $^{13}$CO spectrum over the entire W43 complex (extent defined in Sect. 3.3, see Fig. \ref{f:13co_chanmaps}d). Fig.~\ref{13cospec} presents the $^{13}$CO~1--0  spectra integrated over the entire W43 region, over W43-Main and W43-South. These accumulated spectra allow us to quantify the total flux density at a specific velocity. In Fig.~\ref{13cospec}, there are at least 3 groups of molecular clouds that contribute to the W43 spectrum: in the LSR velocity range $5-15~\kms$, $30-55~\kms$, and $60-120~\kms$. In the larger velocity range, the spectrum is very bright from 80 to 110 $\kms$ and has extensions at $60-80~\kms$ and $110-120~\kms$. 
The spatially integrated emissions in the velocity ranges $5-15~\kms$, $30-55~\kms$, $60-80~\kms$ and $110-120~\kms$ have similar total flux density ($\sim 1 \times 10^{5}$~Jy) and are five times weaker than the main line peak at $80-110~\kms$ ($\sim 3.6 \times 10^{5}$~Jy). We associate the emission in the range $80-110~\kms$ to the bulk of the W43 molecular complex and emissions from other ranges to either (1) diffuse clouds completely unrelated to W43 ($5-15~\kms$ and $30-55~\kms$ range) or (2) clouds which possibly are in the envelope of the W43 complex ($60-80~\kms$ and $110-120~\kms$). The maps integrated for these 5 velocity ranges are given in Figs. \ref{f:13co_chanmaps}a-e.
\begin{figure*}[ht]
\vskip -0.cm
\centering
\includegraphics[angle=-90,width=18.cm]{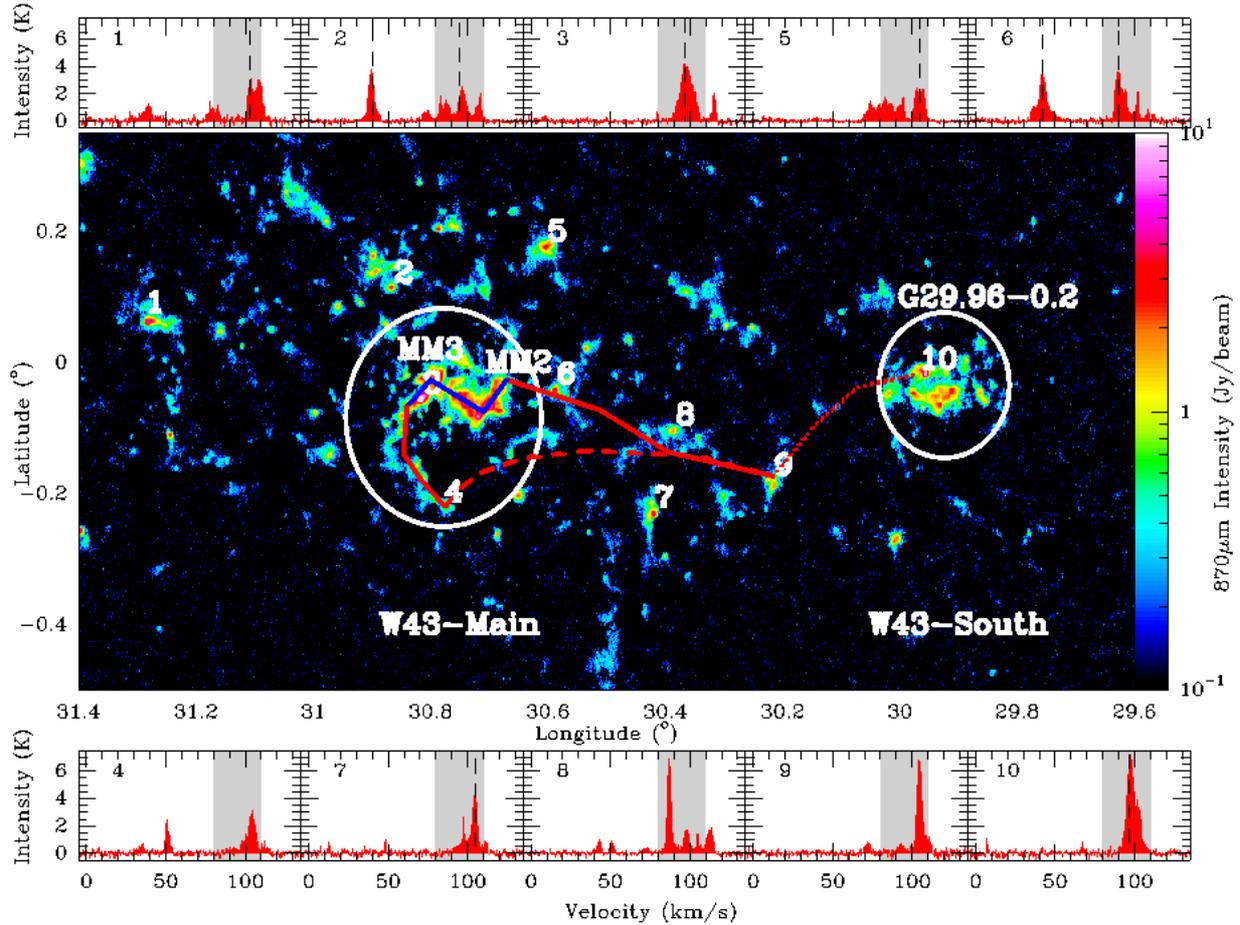} 
\vskip -0.7cm
\caption{870~$\mu$m continuum emission (color scale) and $^{13}$CO~1--0  line spectra observed toward a few bright 870~$\mu$m continuum peaks. In the panels showing spectra taken toward the numbered positions in the image, the main velocity range of W43 ($80-110~\kms$) is indicated with gray filling while dashed lines mark the NH$_{3}$ peak velocities (Wienen et al., in prep.). In the image, the two main bridges identified between W43-Main and South (blue bars) are outlined by red curves which are continuous when they correspond to  $^{13}$CO~1--0  filaments clearly identified by Motte et al. (in prep.) and dashed or dotted for more tentative connections. W43-MM1, W43-MM2 and G29.96-0.2 are indicated as reference points. The white circles locate W43-Main and W43-South, their areas are used to integrate the spectra of Fig.~\ref{13cospec}.}
\label{ATLASGAL}
\vskip -0.5cm
\end{figure*}

\begin{figure}[!h]
\centering
\vskip -0.cm
\includegraphics[angle=0,width=8.cm]{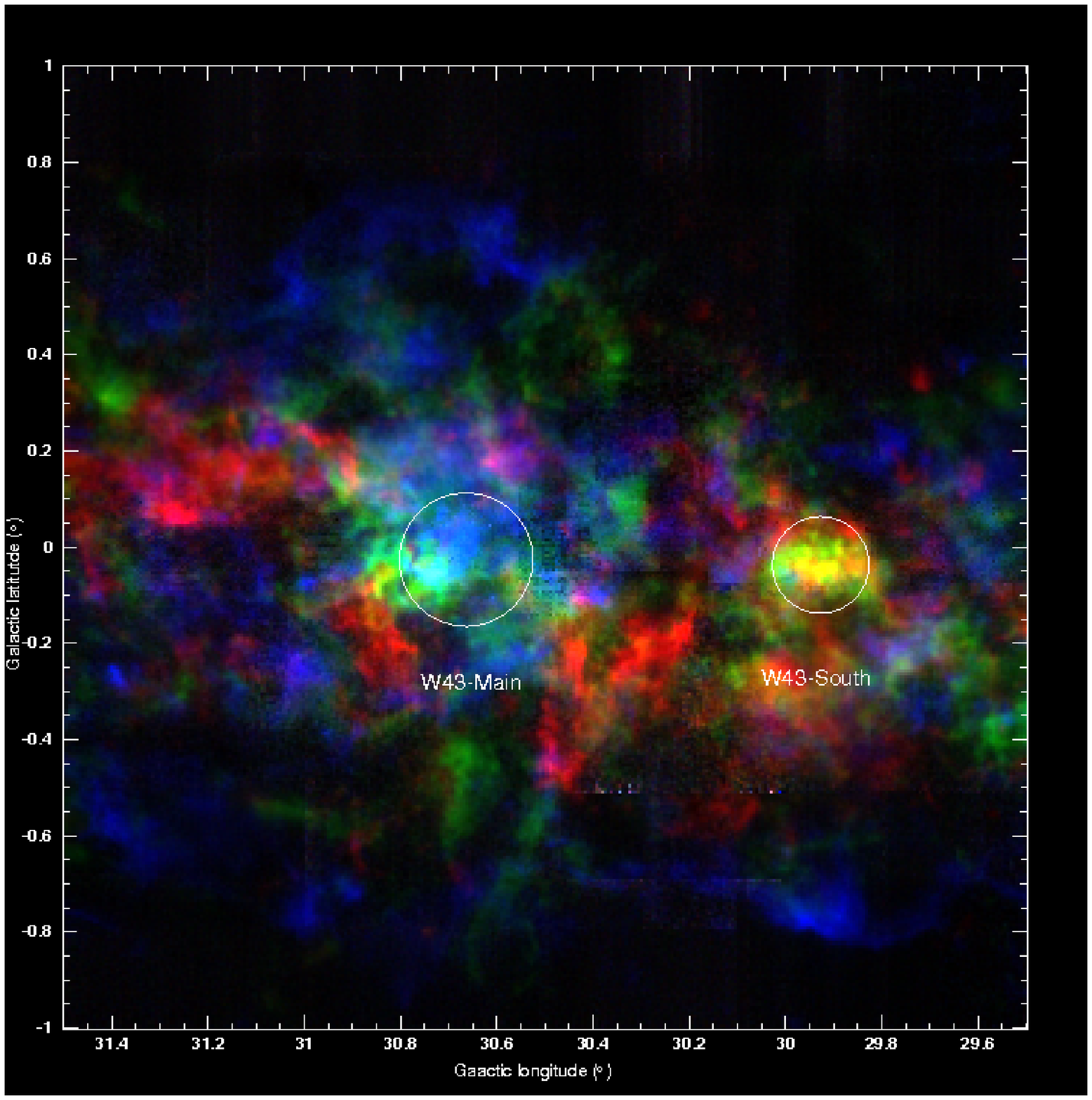}
\caption{Composite 3-color image of the W43 molecular complex emitting in the $^{13}$CO~1--0  line: images are integrated over the velocity ranges $80-90~\kms$ (blue), $90-100~\kms$ (green), $100-110~\kms$ (red). The white circles locate W43-Main and W43-South, their areas are used to integrate the spectra of Fig.~\ref{13cospec}.} 
\label{W43_13CO_3c}
\vskip -1cm
\end{figure}
The main velocity range of W43 is defined to be from 80 to 110 $~\kms$, i.e. with an accumulated flux density of W43 larger than $1/4 \times S^{\mbox{\tiny peak}}\simeq 1 \times 10^{5}$~Jy. This range is tightly associated with the two densest regions, W43-Main and W43-South. The spectrum integrated over the entire W43 is fitted, on the main velocity range of W43 ($80-110~\kms$), by a Gaussian that peaks at $\sim 95.9~\kms$ and has a \emph{FWHM} of $\sim 22.3~\kms$. We first ignored the parts of the $^{13}$CO line which are lower than $80~\kms$  and higher than $110~\kms$ since they are hardly associated with the densest parts of the molecular complex (see Figs.~\ref{f:13co_chanmaps}a-c, e). We however show in Sect.~\ref{singlecomplex} that the high-velocity wing (and possibly part of the low-velocity wing), despite being more than $20~\kms$ away from the median velocity of the W43 molecular complex, consists of lower-density clouds linked in the position-velocity space to the densest parts of W43. Assuming that the dust emission seen at $870\,\mu$m traces the densest parts of the molecular cloud, we check  the coherence of the gas along the velocity axis by comparing $^{13}$CO spectra found at the positions of dust peaks. Figure~\ref{ATLASGAL} shows the dust emission at $870\,\mu$m over the whole W43 complex and identifies two main regions, W43-Main (also named G30.8-0.0) and W43-South (also referred to as G30.0-0.0), and several tens of smaller clumps. Figure~\ref{ATLASGAL} displays the extracted  $^{13}$CO~1--0  spectra at a few bright but randomly selected dust peaks and shows that, except for \#2 and \#6, the spectra have their most prominent peak in the $80-110~\kms$ velocity range. None of the spectra shows a secondary peak at $5-15~\kms$ and only half of them have a secondary peak at $30-55~\kms$. Moreover, LSR velocities measured for these ATLASGAL sources with NH$_3$ observations at Effelsberg all peak in this main velocity range (Wienen et al. in prep.), with a second peak at $30-55~\kms$ for  \#2 and \#6. The $^{13}$CO~1--0  lines integrated over the W43-Main and W43-South regions (as defined in Fig.~\ref{ATLASGAL}) have also their main velocity component within the $\sim80-110~\kms$ range (see Fig.~\ref{13cospec}). We therefore confirm that this velocity range is the most relevant for the W43 complex. Using this main velocity range, we delineate in Fig.~\ref{f:13co_chanmaps}d the W43 molecular complex with a box with longitude from $l=29.6\degr$ to $31.4\degr$ and latitude from $b=-0.5\degr$ to $0.3\degr$. We created a 3-color image of the $^{13}$CO~1--0  emission where blue is $80-90~\kms$, green is $90-100~\kms$, and red is $100-110~\kms$ (see Fig.~\ref{W43_13CO_3c}). W43-Main and W43-South are mainly seen in the green and red ranges, respectively, in agreement with the shape of their integrated line shown in Fig.~\ref{13cospec}.\\
\begin{figure*}[t]
  \centering
  \includegraphics[angle=-90,width=16cm]{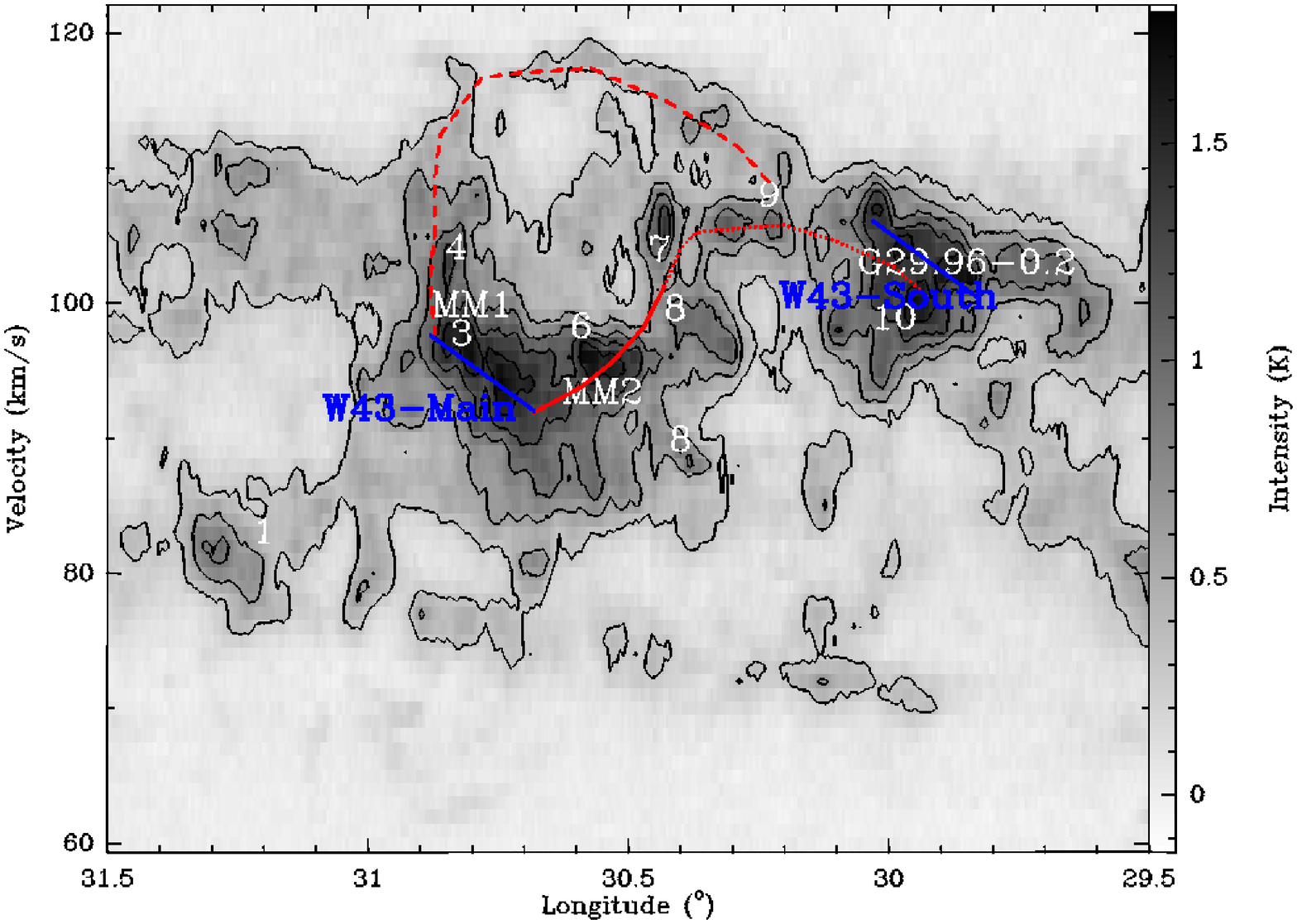} 
\vskip -0.3cm
\caption{The position-velocity diagram of the $^{13}$CO~1--0  line summed over the whole W43 latitude range ($-0.5\degr$ to $0.3\degr$) and plotted against its longitude.  Contours go from to 0.4 to 2.0 by 0.4~K. The two most prominent structures, W43-Main and W43-South, appear to be linked by lower-density clouds. The two main bridges identified between W43-Main and W43-South (blue bars) are outlined by red curves which are continuous when they correspond to filaments clearly identified by Motte et al. (in prep.) and dashed or dotted for weaker clouds connections. The location of a few $870\,\mu$m continuum peaks and reference sources are also indicated.} 
\label{fig:pv}
\end{figure*}

\subsection{W43: a single molecular cloud complex}
\label{singlecomplex}
Most of the cloud structures seen in high-density tracers such as $870\,\mu$m continuum are associated with the W43-Main and W43-South regions. Their spectra peak in velocity ranges close to each other: $\sim 93.8~\kms$ for W43-Main and $\sim 99.1~\kms$ for W43-South, suggesting that they could be connected (see Fig.~\ref{13cospec}). 
\begin{figure*}[t]
\vskip -1.7cm
\centering
\includegraphics[angle=-90,width=21.cm]{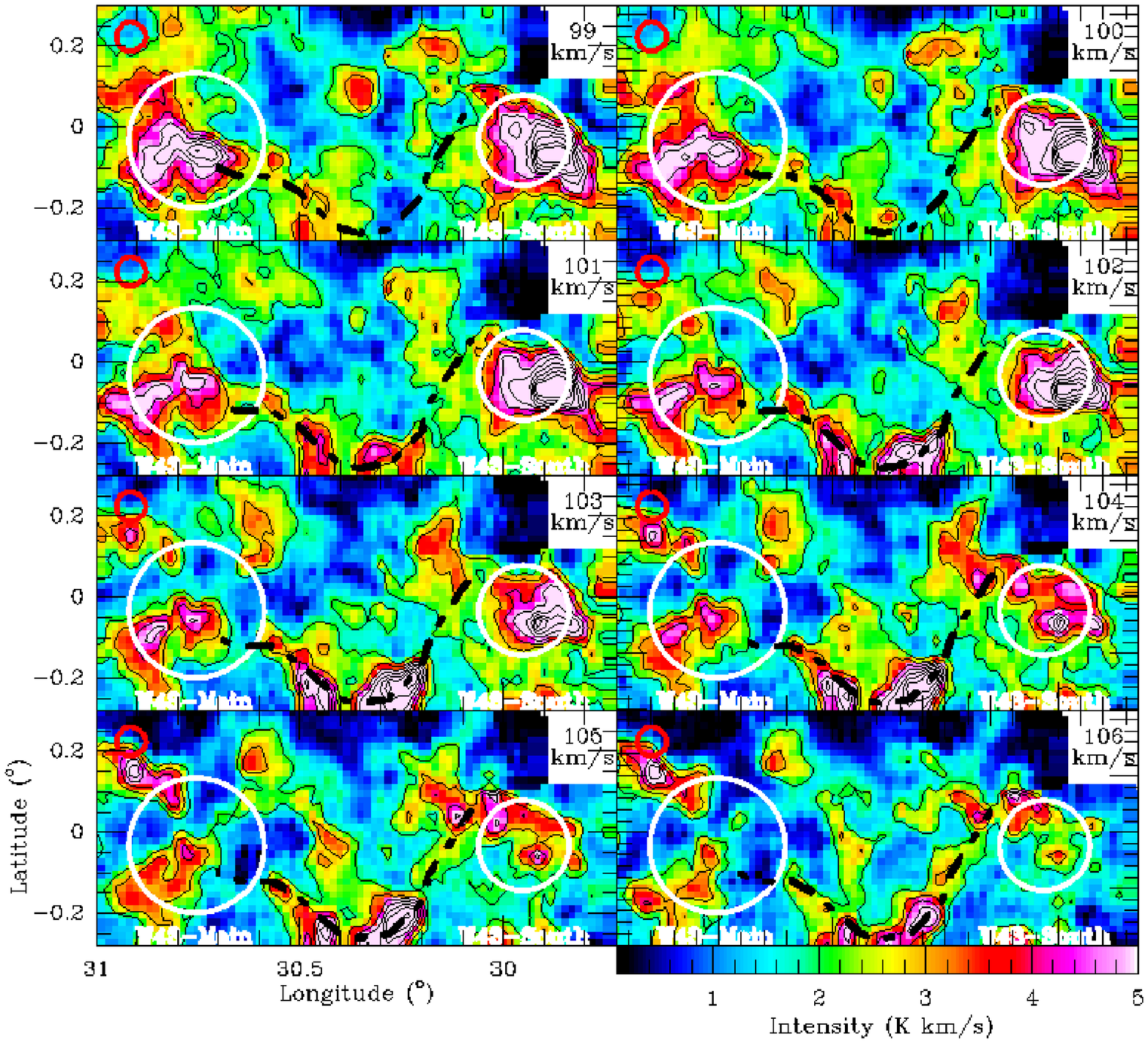} \\

\vskip -0.3cm
\caption{Velocity maps of the inner part of the W43 molecular complex in the $^{12}$CO~2--1  emission obtained with KOSMA, integrated over $\sim$1~$\kms$-wide channels
ranging from $\vlsr=99$ to $106~\kms$. The  W43-Main and W43-South regions are indicated with white circles whose areas are used to integrate the spectra of Fig. \ref{13cospec}. The main molecular bridge linking them is outlined by a dashed curve. The HPBW is plotted in the top-left corner.} 
\label{chan12co}
\end{figure*}

We used  $^{13}$CO~1--0  and $^{12}$CO~2--1  data to test this hypothesis by looking for bridges between W43-Main and W43-South. The position-velocity diagram of Fig.~\ref{fig:pv} displays the $^{13}$CO emission integrated over the full latitude extent of the W43 complex (from $-0.5\degr$ to $0.3\degr$) as a function of the Galactic longitude. Such a diagram argues for W43-Main and W43-South to be  connected by medium- to low-density gas with the peak velocity in the spectra increasing from $\sim 94~\kms$ to $\sim 105~\kms$ and back to $\sim 99~\kms$. This main CO bridge follows a $^{13}$CO filament which connects the southern part of W43-Main (close to W43-MM2) to source \#6, continues south of source \#8 and ends toward source \#9 (see continuous curve in Figs.~\ref{ATLASGAL} and \ref{fig:pv}). The link between source \#9 and source \#10 (equivalent to G29.96-0.2) is more tenuous (see dotted curve in Figs.~\ref{ATLASGAL} and \ref{fig:pv}) but it is obvious in the $^{12}$CO~2--1  channel maps with velocity ranging from $99~\kms$ to $106~\kms$ (see Fig.~\ref{chan12co}). The second CO bridge follows another filament which starts at source \#3 (equivalent to W43-MM1) and connects to source \#4 (see continuous line in Figs.~\ref{ATLASGAL} and \ref{fig:pv}). In Fig.~\ref{fig:pv}, an arc of lower-density material is then linking source \#4 to source \#9 and possibly continues to source \#10, which is G29.96-0.2. Note that this arc of low-density material at velocities $108-120~\kms$ corresponds to the velocity ranges we have excluded from the main velocity range of W43 (Fig.~\ref{13cospec}) and this stream is shown in Fig.~\ref{f:13co_chanmaps}e. Fig.~\ref{fig:pv} also displays a line of diffuse clouds lying from $l=31.3\degr$ and $\vlsr\sim82~\kms$ to $l=29.9\degr$ and $\vlsr\sim 71~\kms$ which may or may not be connected to W43-Main and W43-South. They correspond to the filamentary clouds, shown in Fig.~\ref{f:13co_chanmaps}c, which are mostly away from W43-Main and W43-South. This thus suggests that most of the low-velocity wing of the $^{13}$CO  line in Fig.~\ref{13cospec} has a lower probability to be related to the W43 molecular complex than its high-velocity wing. A similar high-velocity wing feature is found in the W51 region and the authors speculate it is colliding with the W51 main region \citep{carpenter98, 1970A&AS....2..291B}. The fact that W43-Main and W43-South share the same H~{\scriptsize I} envelope (see Sect. 3.4) further strengthens the hypothesis that they belong to the same coherent structure in space and velocity.\\

As we have shown, various evidence supports the hypothesis that W43-Main and W43-South are connected. They are thus probably part of the same molecular cloud structure defined as a GMA. Hereafter we call ``the W43 molecular complex'', the complete ensemble of molecular clouds that covers the area $1.8\degr \times 0.8\degr$ around G30.5-0.1 and the velocity range $80-110~\kms$. With a \emph{FWHM} velocity dispersion of $22~\kms$, W43 is at the upper range of dispersions determined for Galactic and extragalactic GMCs ($5-20~\kms$, \citealt{schneider06}; \citealt{fukui09}). It arises from the fact that the W43 complex consists of an association, with a rather dynamic pattern, of several GMCs which have more typical ($\sim 10~\kms$) dispersions. Note that the remarkably large velocity dispersion of W43 was also noticeable on Fig.~2a of \cite{dame86}. 

\begin{figure*}[t]
  \centering
\includegraphics[width=11 cm,angle=-90]{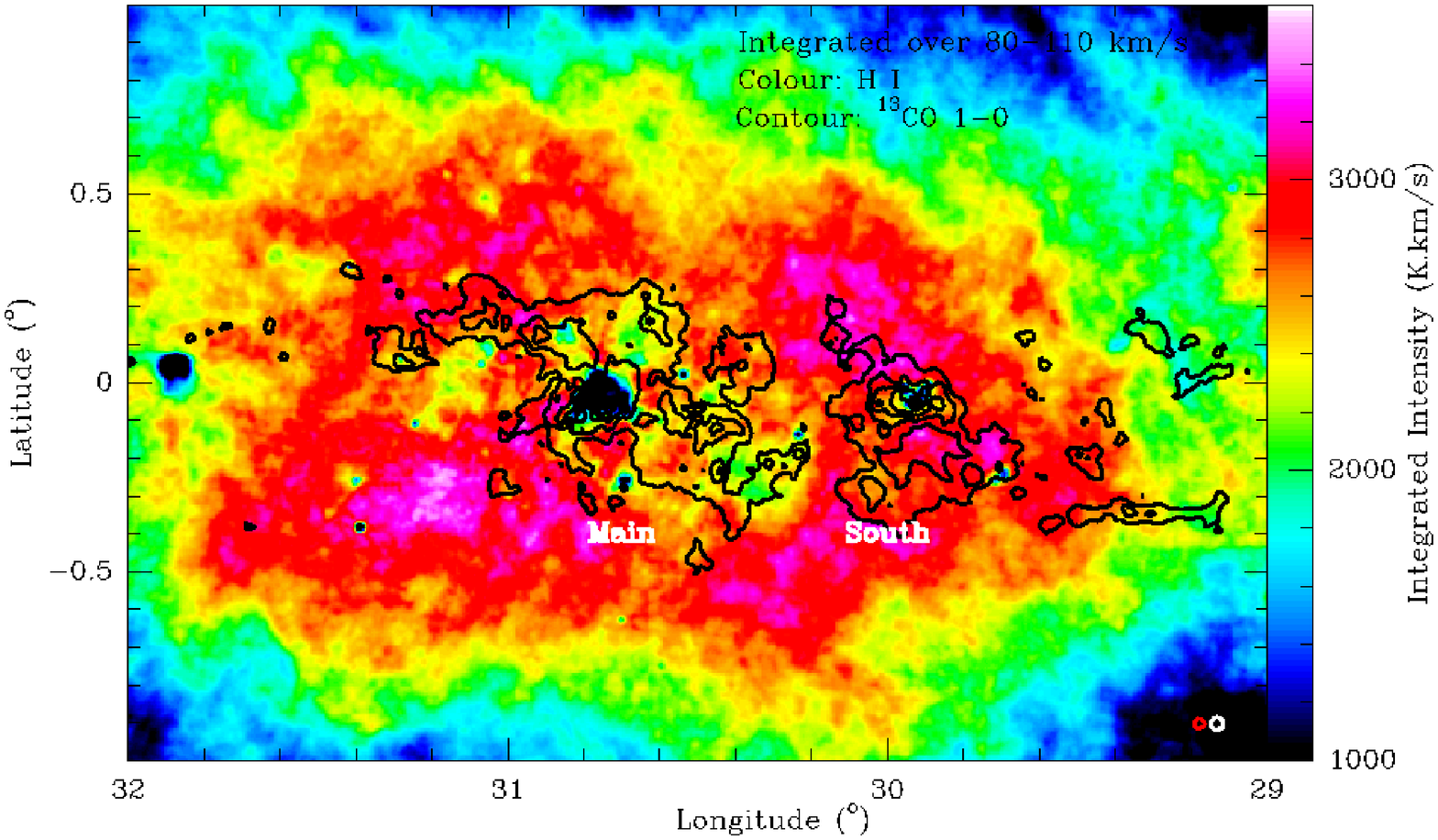}
\vskip -0.5cm
\caption{$^{13}$CO~1--0  intensity map (contours) overlaid on the H~{\scriptsize I} line image of the W43 region. Both lines have been integrated over the $80-110~\kms$ velocity range. Contours go from to 15 to 75 by 10~in unit of K$~\kms$. The H~{\scriptsize I} absorption correlates well with the $^{13}$CO emission of the W43 molecular cloud complex. The HPBWs are plotted in the bottom-left corner for  H~{\scriptsize I} (white) and $^{13}$CO~1--0 (red).}
\label{HI_13CO}
\end{figure*}

\subsection{W43: surrounded by an atomic gas envelope}
\label{hicomplex}

In Fig.~\ref{HI_13CO}, the H~{\scriptsize I} emission associated with the W43 molecular cloud complex, integrated from $80~\kms$ to $110~\kms$, is compared to that of its $^{13}$CO~1--0  emission. Above the $\sim~1500~\K\,\kms$ emission level, the atomic gas surrounds the $^{13}$CO molecular complex. The H~{\scriptsize I} cloud complex of W43 is roughly centered on G30.7-0.05, i.e. $\sim 30$~pc away from the molecular complex center as defined by our analysis of the $^{13}$CO database (G30.5-0.1, see Sect.~\ref{singlecomplex}) and assuming a distance of $\sim 6$~kpc (see Sect.~\ref{sec:location}). The H~{\scriptsize I} map shows strong absorption features where $^{13}$CO is bright, supporting the trend noticed by \cite{liszt95} for W43-Main. In general, these absorption features are caused by self-absorption of the H~{\scriptsize I} line and/or depletion of H~{\scriptsize I} gas that disappears to form molecular gas. The large extent and strength of this self-absorption and/or depletion of the H~{\scriptsize I} emission is not common in the Galactic plane. Together, this suggests that atomic and molecular gas components are intimately linked in the W43 region. We conclude that there is an atomic gas envelope around the W43 molecular complex as well as strong absorption/depletion of atomic gas toward the inner part of the molecular complex. \\
The position-velocity diagram built by \cite{elmegreen87} (see their Figs.~1a-c) for the H~{\scriptsize I} gas in the first Quadrant, suggests a $\sim 60-120~\kms$ velocity range for the W43 H~{\scriptsize I} complex, i.e. a larger range than the one defined in Sect.~\ref{velrange} for the molecular gas. It also displays the 2 other velocity components we identifed in $^{13}$CO ($\sim 5-15~\kms$ and $\sim 30-55~\kms$, see Fig.~\ref{13cospec}). 

\section{Discussion: defining a new and extreme molecular complex}
\label{discussion}

\subsection{W43: the closest molecular cloud of the Galactic Bar}
\label{sec:location}
W43 appears as a coherent complex both in space and velocity and spans the ranges: $l=29.6\degr$ to $31.4\degr$, $b=-0.5\degr$ to $0.3\degr$, and $\vlsr=80-110~\kms$ (see Sect.~\ref{singlecomplex}). We used a standard model for the Galactic rotation \citep{brand93}, with a Galactocentric radius of the Sun of 8.5~kpc to derive the kinematic distance of the W43 molecular cloud complex. This model \citep{2009ApJ...700..137R} assumes that the source $\vlsr$ is only given by the differential rotation in the Galaxy and leads, for sources in the inner Galaxy, to one near and one far kinematic distance. With a median longitude of $l\sim 30.5\degr$ and a median radial velocity of $\vlsr \sim 96~\kms$, we calculated for W43 a near distance of $\sim 5.9$~kpc and a  far distance of $\sim 8.7$~kpc. To resolve the kinematic distance ambiguity, one tries to estimate if most of the cloud material lies in the foreground or background of the source of interest and thus favors, respectively, its far or near distance. Among the most generally-used methods are H~{\scriptsize I} line measurements of H~{\scriptsize II} regions, H~{\scriptsize I} self-absorption features against clouds, and distribution of near-infrared extinction along the line of sight (see e.g. \citealt{anderson09}; \citealt{roman09}; \citealt{schuller09}). In the case of sources in W43, most authors have argued for the near distance: \cite{anderson09} for W43-Main on the basis of H~{\scriptsize I} self-absorption, \cite{pratap99} using both near-infrared extinction and formaldehyde absorption toward the UCH~{\scriptsize II} region G29.96-0.02 in W43-South, \cite{2011A&A...526A.151R} using various techniques on tens of sources of the W43 region. However, \cite{pandian08} using H~{\scriptsize I} self-absorption toward several H~{\scriptsize II} regions of W43-Main favored the far distance. Others simply assumed the average of the near and far distances. Nevertheless, the large-scale absorption of H~{\scriptsize I}, that we investigate more in detail in the forthcoming paper (Motte et al. in prep.) also supports the near-distance of W43.

\begin{figure}[h]
  \centering
  \includegraphics[angle=0,width=8.cm]{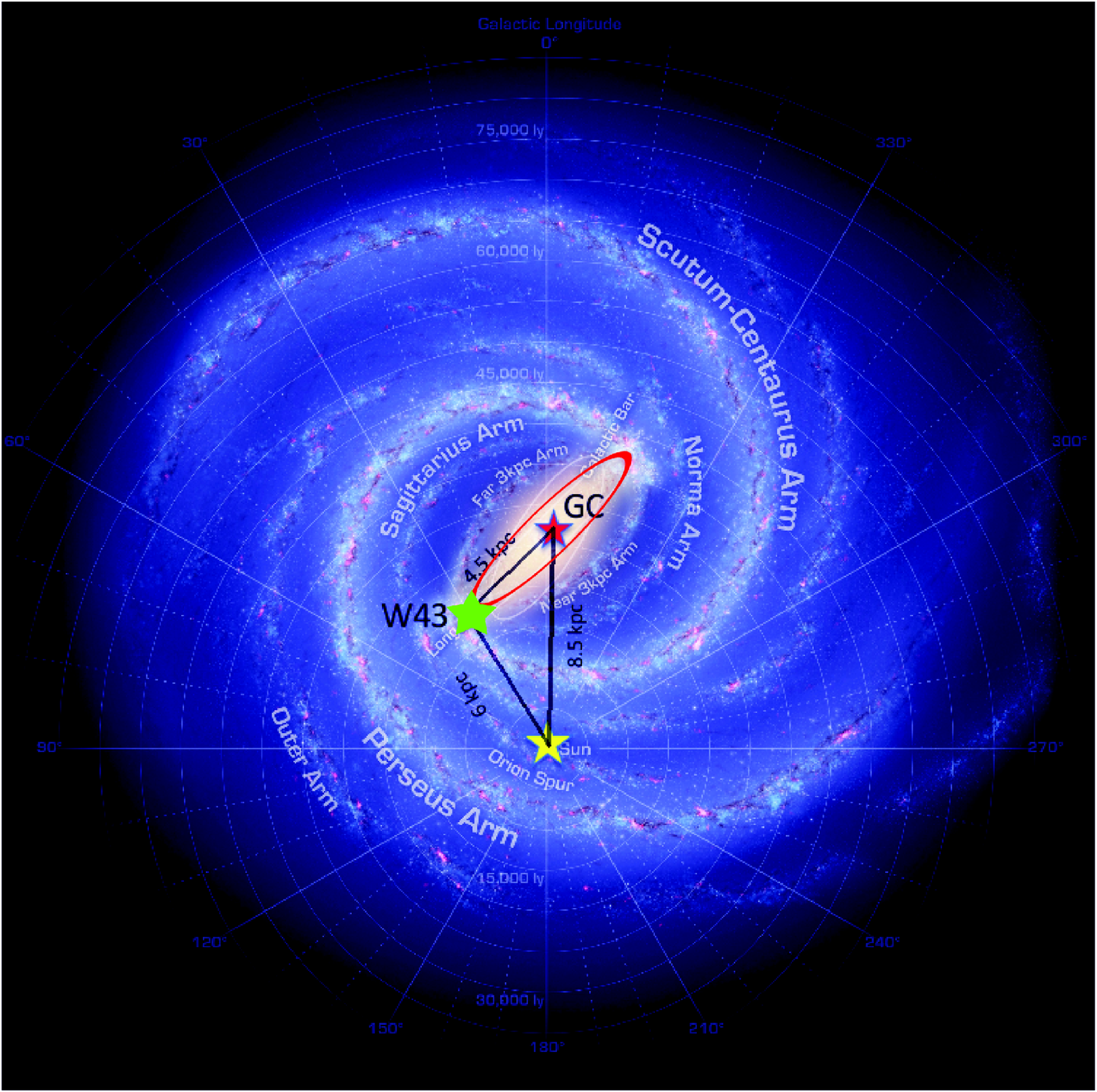} 
\caption{Artist view of the Galaxy seen face-on with the ``long Bar" outlined by a red ellipse \citep{churchwell09}. W43 is located at the expected transition zone between the Bar-dominated region ($R_{GC} <5$~kpc) and the normal Galactic disk.}
\label{galaxyview}
   \end{figure}

The structure of the Milky Way is still a matter of debate but in short it is composed by a disk with spiral arms, a halo, a triaxial bulge, and probably a long bar. Numerical simulations of its gas dynamics assuming a triaxial bulge (boxy inner structure sometimes called the traditional bar) and without long bar would locate W43 in the transition region from the ``lateral arms" and the Scutum-Centaurus arm (e.g. \citealt{rodriguez08}). 
Gas crowding in this region can certainly give rise to the cloud-cloud collisions that may be the origin to the extreme properties of the W43 complex
(N. Rodriguez-Fernandez priv. com.).
Recent observational studies have found evidences of a long in-plane Galactic Bar \citep{HH00, 2005ApJ...630L.149B,2007AJ....133..154L}
that has a radius of 4.4 kpc and is oriented about $44\degr$ to the $\sim8.5$ kpc Sun-Galactic Center line.
From the sketch of this Galactic (long) Bar given in Fig. \ref{galaxyview},
we measure a distance from the Sun to the near-tip of the Bar of $\sim 6$ kpc. The tip of the Bar is the place where the transition from circular to elliptical orbits in the spiral arm and Bar potentials could easily drive high-velocity streams able to collide. 
The gas response to this bar has never been modeled but it will certainly perturb the gas dynamics 
at the position or W43,
contributing to explain the exceptional characteristics of W43 as being a very massive, highly concentrated molecular complex that actively forms (massive) stars (see Table~\ref{table:mass} and Sects.~\ref{totalmass}-\ref{w43atomic}). Other support for our hypothesis comes from the fact that extragalactic studies often observe molecular clouds and star-forming regions with extreme properties at the ends of galactic bars (e.g. M~83, NGC~1300, see \citealt{1997A&A...326..449M} and references therein).
Note also that the determination of the distance through kinematic distance has limitations due to the fact that the true rotation pattern of our Galaxy and especially at the connection point of Galactic arms and the Bar, may significantly deviate from the theoretical assumptions of axially symmetric and circular orbits \citep{russeil03, 2009ApJ...700..137R}. A $\sim 10~\kms$ deviation due to streaming motions is typical in Galactic arms and would give kinematic distances of $5.9_{-0.7}^{+1.2}$~kpc and $8.7_{-1.3}^{+0.6}$~kpc. Such a velocity deviation is in fact smaller than the velocity dispersion measured in molecular complexes such as W43 and Cygnus~X (see Table~\ref{table:mass}). Our study therefore suggests that the W43 molecular complex could be located at $\sim 6$~kpc from the Sun and at the connecting point of the Scutum-Centaurus arm and the Galactic Bar. The stellar distance of clusters or the parallax of masers excited by H~{\scriptsize II} regions associated with the W43 molecular cloud could help us to check this assumption. Up until now, the very few attempts to use the complementary spectro-photometric method on the WR-OB cluster of W43-Main and on the 2 OB stars within G29.96-0.02 have not given convincing results (d$\sim~$4.3 kpc, \citealt{pratap99}), probably because the stars are extremely young and highly extinct. It is unfortunately not yet possible to measure the parallax distance of W43 masers since they are rather weak.

\begin{table*}
\caption{Global characteristics of the W43 molecular and star-forming complex in comparison with other star-forming complexes}
\label{table:mass}
\centering
\begin{tabular}{l l l l l l l l l}
\hline
\hline
Region 		& Assumed  & Equivalent Diameter	& Velocity Range	& Total H$_2$ Mass	 & Clouds Mass	& Dense Clumps Mass	& $8~\,\mu$m Luminosity\\ 
			& Distance & From $^{13}$CO      &  From $^{13}$CO     	         &  From $^{12}$CO        &  From $^{13}$CO        &  From submm $^{\mbox{~a}}$                               & From 8~$\mu$m $^{\mbox{~b}}$\\
			&~~~[kpc] & ~~~~~~~~[pc]		& ~~~~~[$\kms$]			& ~~~~~[$\msun$]                 & ~~~~~[$\msun$]                        & ~~~~~~~[$\msun , \% $]           & ~~~~~~~[$\lsun$] \\
\hline
W43			& ~~6.0   		   & $\sim 140$			   & ~~~80 to 110		           & ~~~$7.1 \times10^6$                         & ~~~$4.2\times10^6$                         & $8.4\times10^5$ ,  12\%              &  $1.60\times10^7$\\
Cygnus~X	&$~~1.7^{\mbox{~c}}$ &  $\sim 160$$^{\mbox{~c}}$	   & ~~-10 to 20$^{\mbox{~c}}$	  & ~~~$5.0\times10^6$$^{\mbox{~c}}$           &  ~~~$8.0\times10^5$$^{\mbox{~c}}$         & $6.0\times10^4$ , 1\% $^{\mbox{~d}}$ &$ 6.57\times10^6$$^{\mbox{~e}}$\\
W51		        &  ~~7.0 	   & $\sim 100^{\mbox{~f}}$           & ~~~56 to 65$^{\mbox{~f}}$          & $\sim1.2\times10^6$$^{\mbox{~f}}$ 						      &   ...      &      $5.5 \times10^5$                                     & ...\\
CMZ			& ~~8.5    & $\sim 350^{\mbox{~g}}$ 	   & -225 to 225$^{\mbox{~g}}$	  & $\sim 3.0 \times10^8$$^{\mbox{~g}}$  & $\sim 2.0 \times10^7$$^{\mbox{~g}}$ & $4.0 \times10^6$                               &...\\
W49		         & 11.4$^{\mbox{~h}}$  & ~~$\sim 45^{\mbox{~h}}$ 	   & ~~~~-5 to 25$^{\mbox{~h}}$ 	  &   ...                                                     & $\sim5.0\times10^5$$^{\mbox{~h}}$       &         $2.0 \times10^5$                             &...\\
\hline
\end{tabular}
\begin{flushleft}
Notes:
(a) The dense clumps mass is calculated from the ATLASGAL 870~$\mu$m images and MAMBO2 1.3 mm for Cygnus X, on which $\geqslant$~5 pc cloud structures have been filtered (more precisely, 5.6 pc for W43, 5 pc for Cygnus X, 6.5 pc for W51, 4.0 pc for CMZ, and 5.3 pc for W49) and using Eq. \ref{eq:mclump}. The percentage of dense clumps formed out of the total H$_2$ gas (taken from column 5) is estimated for the regions where the total H$_2$  mass is more reliable; 
(b) The $8~\,\mu$m luminosity is measured from band 4 of \emph{Spitzer} which has a bandwidth of $2.93~\,\mu$m centered at  $7.91~\,\mu$m;
(c) From \cite{schneider06}; 
(d) From \cite{motte07} at 1.2mm; 
(e) From \cite{hora09}; 
(f) From \cite{carpenter98}; 
(g) From \cite{1998A&A...331..959D} using C$^{18}$O; 
(h) From \cite{simon01}. 
\end{flushleft}
\end{table*}

\subsection{The total mass of the W43 molecular cloud complex}
\label{totalmass}
We determined the total H$_2$ mass of the W43 molecular complex from the $^{12}$CO~1--0  data of the CfA survey \citep{dame01}. We measured the total intensity of $^{12}$CO~1--0  over the full area of the W43 complex ($A=1.8^\circ\times0.8^\circ$) and over the velocity range of $80-110~\kms$. We then calculate the total gas mass using the conversion factor $N_{\mbox{\tiny H2}} = 2.75 \times 10^{20}\, \times\, I(^{12}\mbox{CO})$ \citep{bloemen86} and the following equation: 
\begin{equation}
M_{\mbox{\tiny total}}= N_{\mbox{\tiny H2}} \times\, A\, \times\, m_{\mbox{\tiny H2}} = 7.1 \times 10^6~\msun \, \times \,\left(\frac{d}{\mbox{6~kpc}} \right)^2
\end{equation}
The mass estimated with this empirical equation is rather uncertain since some clouds may still lie along the same line of sight, unrelated to W43, but it gives an upper limit to the total mass of W43. The computed mass is in the range of values one finds for large molecular cloud structures such as complexes and GMAs (e.g. \citealt{kuno95}).
We also calculated the molecular mass of dense clouds in W43 from the $^{13}$CO~1--0  and $^{12}$CO~2--1  data. We used the generally adopted scheme which assumes a uniform medium (filling factor of 1), and that the emission is optically thick in $^{12}$CO and optically thin in $^{13}$CO (e.g. \citealt{rohlfs00, schneider06}). 
The  excitation temperature is first measured from the $^{12}$CO~2--1  main beam temperature and assumed to apply to the $^{13}$CO~1--0  transition. We determined $T_{\mbox{\tiny ex}}$ for each pixel of the $^{12}$CO map and showed that it ranges from 5 to 18~K with a mean value of $\sim 10$~K. Under the LTE assumption, the column density of $^{13}$CO can then be estimated from its intensity with the equation 
$N(^{13}\mbox{CO})= \frac{3.0 \times 10^4~\mbox{\tiny cm}^{-2}}{1-(e^{-5.3~ K/T_{ex}} )} \int \tmb(^{13}\mbox{CO})~dV$. The total column density of $^{13}$CO is finally converted into H$_{2}$ column density by multiplying by a factor of $4.7 \times 10^5$ \citep{rohlfs00}. We assumed an excitation temperature of $\sim 10$~K for each point and estimate the mass from the $^{13}$CO intensity integrated over the full extent of the W43 complex: 

\begin{equation}
M_{\mbox{\tiny clouds}} \sim 4.2 \times 10^6~\msun\, \times \,\left(\frac{d}{\mbox{6~kpc}} \right)^2.
\end{equation}
The difference between $M_{\mbox{\tiny total}}$ and $M_{\mbox{\tiny clouds}}$ illustrates the uncertainty inherent to the mass estimation of CO clouds. These values are in good agreement with the virial masses measured by \cite{mooney95} for the two main clouds W43-Main and W43-South and summing up to $\sim 4 \times 10^6~\msun$. Note that this $M_{\mbox{\tiny clouds}}$ value is slightly underestimated since some of the $^{13}$CO  clouds are optically thick.
An estimate of the virial mass of W43 was performed using the Gaussian fit to the $^{13}$CO line shown in Fig.~\ref{13cospec} ($\Delta V\sim 22~\kms$ and thus $\sigma=\sqrt{3/(8 \ln2)} \times \Delta V\sim 16~\kms$) and the diameter given in Table~\ref{table:mass} is equivalent to $R\sim 70$~pc: 
\begin{equation}
M_{\mbox{\tiny vir}} = \frac{5 \sigma^2 \times R}{G} = 2.1 \times 10^7~\msun\, \times \,\frac{d}{\mbox{6~kpc}} .
\end{equation}
The ratio $M_{\mbox{\tiny total}}/M_{\mbox{\tiny vir}}\sim 0.35$ suggests that the entire W43 molecular complex is probably in gravitational virial equilibrium \citep{pound93}. If we only consider the molecular clouds securely associated with W43-Main (from l=30.2$\degr$ to 31$\degr$), we estimate $M_{\mbox{\tiny total}}= 5.0 \times 10^6 \msun$, $M_{\mbox{\tiny clouds}}= 3.0 \times 10^6 \msun$, $M_{\mbox{\tiny vir}}=  1.0 \times 10^7\msun$ leading to $M_{\mbox{\tiny total}}/M_{\mbox{\tiny vir}}\sim 0.5$. These results show that both the W43-Main complex and the entire W43 GMA should be in gravitational virial equilibrium, like most of the observed molecular clouds or complexes (e.g. \citealt{heyer01}). They further support the idea that the W43 GMA is indeed a single molecular cloud entity bounded by self-gravity and that its individual molecular clouds could even be collapsing.      

\subsection{The mass of compact clumps in W43}
\label{clumpsmass}
The ATLASGAL image of W43 (see Fig. 4) probes the distribution of relatively large (up to $\sim$10~pc) cloud structures through their dust continuum emission at $870\,\mu$m. In order to trace the mass of clumps with size $< 5$~pc, we filtered out the emission on large scales by applying the wavelet transform technique of \cite{starck06} as it was used by \cite{motte07}. If we assume that the $870\,\mu$m emission is optically thin and mostly consists of thermal dust emission, the total gas $+$ dust mass of the W43 clumps is simply proportional to its total flux, $S_{870}$,
integrated over the $1.8^\circ \times 0.8^\circ$ area defined in Sect.~\ref{results}. We used an equation similar to Eq.~1 of \cite{motte03} and estimated a mass for W43 clumps of
\begin{equation}
\label{eq:mclump}
 M_{\mbox{\tiny clumps}} = 8.4\times 10^5~\msun \left( \frac{e^{\frac{16.58}{\tdust}}-1}{e^{\frac{16.58}{\mbox{\tiny 20 K}}}-1} \right) 
  \left( \frac{\kmm}{0.015~\cmg} \right)^{-1} \left(\frac{d}{\mbox{6~kpc}} \right)^{2}. 
 \end{equation}
We assumed a dust temperature of 20~K, which is typical of compact cloud structures in high-mass star-forming regions and is in good agreement with the $\tdust$ values measured by \cite{motte03} on a few W43 clumps and by \cite{molinari10} on the W43 complex. We also assumed a dust emissivity of $\kmm=0.015~\cmg$, which is intermediate between that recommended for protostellar envelopes and prestellar cores \citep{Ossenkopf94}. The mass estimate of each clump is uncertain by a factor of $\sim 2$ due to our poor knowledge of its dust temperature and emissivity but over the complete cloud the average uncertainty should be lower. Since dust emissivity may increase with the metallicity of the medium and thus be larger at a Galactocentric distance of 4~kpc (W43) than at 8~kpc (solar neighborhood), the mass given above could be overestimated by up to a factor of 1.4 (see \citealt{mooney95} and references therein). Note that taking the metallicity into account would also decrease the mass estimates of 
 $M_{\mbox{\tiny total}}$ and  $M_{\mbox{\tiny clouds}}$.\\

\subsection{W43 compared to other prominent cloud complexes}
Through the Galactic Plane, the most prominent cloud complexes are relatively nearby (at 1--3 kpc from the Sun) like M16-M17, NGC 6334-6357, Cygnus~X, NGC~7538, W48, or W3. Further away, one has found extreme star-forming regions such as W43, W49, or W51 and of course the Central Molecular Zone (CMZ). A complete census and characterization of massive molecular complexes in our Galaxy is underway in the framework of the ATLASGAL and Hi-GAL surveys (see first study by \citealt{schuller09} and \citealt{molinari10}). Given our present view, we however believe that W43 will remain one of the most extreme molecular complexes in the Milky Way that is also relatively near to the Sun.\\
The W43 molecular complex is located more than three times further away from the Sun than Cygnus~X but has a similar linear size. The total mass of H$_2$ gas in W43 is very high, even slightly higher than in the Cygnus~X molecular complex (see Table~\ref{table:mass}) which is already among the most massive complexes of our Galaxy. This trend is larger when we compare, in W43 and Cygnus~X, the mass accumulated within clouds identified in $^{13}$CO and clumps ($< 5$ pc cloud structures): it is $\sim 5$ and almost 15 times larger in W43. Interestingly enough, the clumps-to-total mass ratio found in W43 is $\sim10\%$, definitively much larger than that found in Cygnus~X ($\sim 1\%$, \citealt{motte07}). $\sim20\%$ of the $^{13}$CO cloud mass in W43 is located in high-density clumps, which is similar to that observed for extreme star-forming regions such as W49, W51 and the CMZ, despite the $^{13}$CO cloud mass being measured differently. W43 is thus a massive molecular complex ($7.1 \times 10^6~\msun$ within a linear diameter of $\sim 140$~pc) which is exceptional in terms of the concentration of its gas into $^{13}$CO cloud ($4.2 \times 10^6~\msun$) and compact cloud structures ($8.4 \times 10^5~\msun$ within 5~pc dense clumps). 
The velocity dispersion determined by a Gaussian fit ($\Delta V\sim 22~\kms$) for W43 is so large that most of the previous studies separated the region into two complexes, located over the same range of Galactic coordinates but at different velocities and with more ordinary internal motions (at $\vlsr \sim 80~\kms$ and $\sim 100~\kms$ with $\Delta V\sim 15~\kms$, e.g. \citealt{dame86}). We have shown that W43 can be defined as a single and coherent molecular complex (see Sects.~\ref{singlecomplex}-\ref{hicomplex}), which consists of several velocity streams/density filaments associated with the densest parts of W43. It is difficult to compare the velocity dispersions of W43 to other cloud complexes listed in Table 2 since the Gaussian fits are generally not computed in the literature. However, considering the velocity spans of those clouds, the total velocity dispersion of W43 seems significantly larger than that of Cygnus X, W49, and W51.

\subsection{The H~{\scriptsize I} envelope of W43 and the W43 (H~{\scriptsize I} $+$ H${_2}$) cloud} 
\label{w43atomic}

The H~{\scriptsize I} gas is enveloping the W43 molecular cloud and is thus covering a larger area and possibly a larger velocity range than the $^{13}$CO cloud. From Fig.~\ref{HI_13CO}, we estimate that it covers the whole image, i.e. $3^\circ \times 2^\circ$ or a mean diameter of $\sim 290$~pc 
and the position-velocity diagram of \cite{elmegreen87} gives a velocity range for the H~{\scriptsize I} gas that could be twice as large as that of molecular complexes: up to $\sim 60-120~\kms$ . In contrast to $^{13}$CO, the confusion along the line of sight is very large at a longitude close to $l\sim 30^\circ$ since the intensity measured on the total velocity range of the VGPS ($-120$ to 170~\kms) is $\sim 4.4$ times larger than the one measured by limiting the velocity range to $80-110~\kms$. For a meaningful comparison of the atomic and molecular gas, we only consider H~{\scriptsize I}  emission within that range. 

We determined the column density of H~{\scriptsize I} assuming that the W43 H~{\scriptsize I} cloud is uniform and optically thin and using its integrated intensity over the velocity range $80-110~\kms$ in the equation $N_{\mbox{\tiny H~{\scriptsize I}}} =1.82 \times 10^{18} ~\mbox{cm}^{-2} \int \tmb(\mbox{H~{\scriptsize I}})~\mbox{dV}$ \citep{spitzer78}. The column density map is then integrated over the $3^\circ \times 2^\circ$ area and gives $M_{\mbox{\tiny H~{\scriptsize I}}} \sim 3.2 \times 10^6~\msun$, which agrees with the value calculated by \cite{elmegreen87} under the same assumptions. Approximately 20\% of the H~{\scriptsize I} image of Fig.~\ref{HI_13CO} displays absorption features. If they correspond to self-absorption, its high-level (by factors of $\sim 1.5$ in average and up to $> 3$ in W43-Main) could lead us to underestimate the total H~{\scriptsize I} mass by up to 50\%, i.e. $ \sim 6.4 \times 10^6~\msun$. We therefore estimate of the H~{\scriptsize I} gas mass to be:
\begin{equation}
M_{\mbox{\tiny H~{\scriptsize I}}} \sim 3.2-6.4 \times 10^6~\msun\, \times \,\left(\frac{d}{\mbox{\tiny6~kpc}} \right)^2.
\end{equation}

The estimates of Sects.~\ref{totalmass} and \ref{clumpsmass} give a total mass for the H~{\scriptsize I} $+$ H${_2}$ cloud of $\sim 1-1.2\times 10^7~\msun$. The molecular mass fraction of W43 is roughly estimated to be $\sim 65\%$.
This agrees with the tendency to have a large molecular mass fraction ($\sim 70\%$) in the Galactic center and a small one ($\sim 5\%$) in the outer Galaxy \citep{elmegreen87}. This may suggest a stronger link between the H~{\scriptsize I} and H${_2}$ gas components in W43 as well as a faster formation of molecular clouds than in the outer Galaxy. The special location of W43 may have helped to accumulate H{\scriptsize~I} gas from different structures (arm, halo, Bar) to become a large potential well and form molecular clouds very fast and efficiently. When massive stars are formed, the molecular gas is photo-dissociated back to become H{\scriptsize~I} again. These two sources of H{\scriptsize~I} create the envelope-like emission morphology seen in Fig.~\ref{HI_13CO}. 

\subsection{W43: An exceptional star-forming region}
 
The W43 molecular complex is associated with rich clusters of massive stars as well as more recent episodes of star formation given the existence of high-mass protostars and CH {\scriptsize II} regions. 
W43-Main is well known for its giant H~{\scriptsize II}  region emitting 10$^{51}$ Lyman continuum photons s$^{-1}$ and having a far-infrared continuum luminosity of $\sim 3.5\times 10^{6}~ \lsun$ (\citealt{smith78}). It hosts an ionizing source which was discovered in near-infrared images by \cite{Lester85} and confirmed by \cite{blum99} as a cluster of WR and OB main-sequence stars. The large luminosity and ionizing flux of the W43-Main star-forming region is comparable to that of the very massive star cluster NGC~3603 or M17, which suggests that the central cluster in W43 contains a large number of as yet undetected massive stars. The submillimeter continuum and HCO$^{+}$(3--2) survey of  \cite{motte03} revealed $\sim 50$ massive ($20-3600~\msun$) clumps probably forming high-mass stars thus suggesting that the molecular cloud is  undergoing a second remarkably efficient episode of high-mass star formation (\emph{SFE}$\sim 25\%/10 ^6$~yr). The W43-South region also hosts the well-studied CH~{\scriptsize II} region/hot molecular core G29.96-0.02 which has a luminosity of $\sim 4.4\times 10^5~\lsun$ (\citealt{wood89}; \citealt{cesaroni94}; \citealt{cesaroni98}). This CH~{\scriptsize II} is excited by an O5-O8 star (\citealt{watson97}) and is associated with an embedded cluster containing massive stars located at its rim \citep{pratap99}. 

The $8\,\mu$m flux integrated over the W43 molecular complex gives a rough estimate of the recent (high-mass) star formation activity. Indeed, the \emph{Spitzer}/IRAC band 4 is dominated by the emission of PAH particles which is tracing the current UV field and thus the number of young OB stars and H~{\scriptsize II} regions (e.g. \citealt{peeters04}). We calculated the $8\,\mu$m luminosity of W43 from the $8\,\mu$m flux to allow comparison with other star-forming regions (see Table~\ref{table:mass} and Sect.~\ref{Q1}) based on the following equation:\\
 \begin{equation}
 \label{eq:flux}
L
=2.74\times 10^{-10} \cdot \frac{F}{\left[\mbox{MJy/sr}\right]} \cdot  \frac{A}{\left[\arcsec^{2}\right]} \cdot  \frac{d^{2}}{\left[\mbox{pc}^{2}\right]}~ \left[\lsun\right]
 \end{equation}
                
Since W43 lies in the inner part of the Galactic plane, its $8\, \mu$m luminosity should be corrected for emission arising from other regions along the same line of sight. We estimated this background and foreground emission in the W43 region by measuring the $8\, \mu$m fluxes just around the molecular complex where there are no dominant emissions from compact sources. The contaminating emission accounts for about 1/3 of the total flux. Table~\ref{table:mass} reports the $8\, \mu$m luminosity corrected for the infrared background+foreground for W43 and the uncorrected $8\,\mu$m luminosity for Cygnus~X. We show that the $8\,\mu$m luminosity and thus the star formation activity of W43 could be almost 3.5 times larger than that of  
Cygnus~X. This is remarkable since Cygnus~X is recognized to be one of the most active star-forming complexes of our Galaxy \citep{hora09}. We estimate the ``current" star formation rate(\emph{SFR}) of W43 using the $8~\,\mu$m luminosity measured here and equations derived by \cite{wu05} from the \emph{Spitzer} imaging of galaxies and the \emph{SFR} formula of \cite{kennicutt98}:
\begin{equation}
\mbox{\emph{SFR}}_{8~\micron}=\frac{\nu L_{\nu}[8 \micron]}{1.57 \times 10^9~ \lsun}\sim 0.01~\msun\, \mbox{yr}^{-1}\, \times \,\left(\frac{d}{\mbox{\tiny6~kpc}} \right)^2.
\end{equation}

The $8\, \mu$m \emph{SFR} traces the young and already-formed stars, but it does not hold for the \emph{SFR} in the future. The latter can be estimated from the molecular gas content in W43. Assuming that the W43 molecular complex, with a mass of $\sim 7.1 \times 10^6~\msun$, is forming stars with \emph{SFE} of a few percent (\emph{SFE}$\,\sim 1\%-3\%$, \citealt{silk97}), we expect it to form $\sim 1.4\times 10^5~\msun$ of stars. With a typical lifetime of high-mass star-forming regions of $\sim 1-3 \times 10^6$~yr \citep{roman09}, in agreement with scenarios of cloud formation from colliding flows (\citealt{heitsch08}; \citealt{hennebelle08}; \citealt{vazquez07}), this estimate would yield a ``future" \emph{SFR} of:
\begin{equation}
\mbox{\emph{SFR}}_{\mbox{\tiny CO}}= 0.05-0.14~\msun\, \mbox{yr}^{-1}\, \times \,\left(\frac{M_{\mbox{total}}}{7.1\times 10^6~\msun} \right) \times \left(\frac{\mbox{\emph{SFE}}}{2\%} \right).
\end{equation}
The lower value is $5$ times larger than the ``current" SFR obtained from the $8\,\mu$m luminosity. The upper value is even larger ($14$ times larger) and agrees with the estimate made by \cite{motte03} for W43-Main only. This result confirms that W43 will be very efficient in forming stars, $\sim10$ times more efficient than it has already been in the past, despite the presence of the young ($\sim 10^6$~yr) cluster of WR-OB stars in W43-Main. We may be witnessing the formation of new starburst clusters in the W43 region. With a \emph{SFR} of $\sim 0.1~\msun\, \mbox{yr}^{-1}$ and a lifetime of $\sim 2\times 10^6$~yr, we should expect $\sim 2\times 10^5~\msun$ of stars to form in the W43 molecular cloud, among which there should be about $2.4\times 10^4~ \msun$ (12\% of total stellar mass) of stars with mass larger than $8~\msun$ using the Initial Mass Function (IMF) of \cite{kroupa01}. If all of this mass would convert into stars with $8~\msun$ or $50~\msun$, the total number of such stars would be $\sim 3000$ or $\sim 500$, respectively.

\section{Conclusions}

In the framework of the ATLASGAL survey of star formation, we here identify a new molecular cloud complex located around G30.5-0.1, including W43-Main and W43-South. We used a large database tracing diffuse atomic gas (H~{\scriptsize I} emission at 21~cm from the VGPS survey), low- to medium-density molecular gas ($^{12}$CO~1--0  and $^{13}$CO~1--0  emission from the CfA and GRS surveys), high-density molecular gas ($870~\mu$m continuum emission from the ATLASGAL survey), and star formation activity ($8~\mu$m from the GLIMPSE survey) and obtained KOSMA observations of the $^{12}$CO~2--1 , 3--2 line emission. Our main findings can be summarized as follows:

\begin{enumerate}
\item From the detailed 3D (space-space-velocity) analysis of the molecular  tracers ($^{13}$CO~1--0  data cubes of the GRS and $^{12}$CO~2--1 , 3--2 data cubes of KOSMA) through the region, we identified W43 as a coherent complex of molecular clouds. It covers a spatial extent of $\sim 140$~pc and a velocity range of $\sim 22.3~\kms$ \emph{FWHM} and spans the ranges $l=29.6\degr$ to $31.4\degr$, $b=-0.5\degr$ to $0.3\degr$, and at least $\vlsr=80-110~\kms$. These values show that W43 is a large complex of clouds which has a wide velocity dispersion.
\item The analysis of the atomic gas data cube from the VGPS argues for the W43 molecular complex being surrounded by an atomic gas envelope of larger diameter, $\sim 290$~pc, and the same or larger velocity range, up to $\sim 60~\kms$. 
\item The distance we estimated for the W43 complex is $\sim 6$~kpc from the Sun. It locates it at the meeting point of the Scutum-Centaurus Galactic arm and the Bar. This point is a very dynamic region of our Galaxy since it coincides with the place where the transition from circular to elliptical orbits in the spiral arm and bar potentials could easily drive high-velocity streams able to collide. 
\item We measured the total mass of the W43 molecular complex to be $M_{\mbox{\tiny total}}\sim 7.1 \times 10^6~\msun$ and the mass contained in dense $870~\mu$m  clumps ($<5$~pc dense cloud structures) to be $M_{\mbox{\tiny clumps}}\sim 8.4\times 10^5~\msun$. When compared to Cygnus~X and the Central Molecular Zone, these values make W43 a massive and concentrated molecular cloud complex. These findings are in agreement with W43 being in the region sometimes called the Molecular Ring and known to be particularly rich in terms of molecular clouds and star formation activity. 
\item We estimated the \emph{SFR} of W43 using 1) the $8\,\mu$m luminosity measured by \emph{Spitzer} in W43 and 2) the mass of W43, the classical \emph{SFE} and cloud lifetime, obtained an increase from \emph{SFR}$\, \sim 0.01~\msun\, \mbox{yr}^{-1}$ $\sim 10^6$~yr ago to $0.1~\msun\, \mbox{yr}^{-1}$ in the near-future. We may be witnessing the formation of new starburst clusters in the W43 region. This result generalizes the study of \cite{motte03} which was dedicated to W43-Main only.
\item We compared  the global properties (mass, density, dynamic state, star-formation activity) of the W43 molecular cloud and star-forming complex to other prominent regions and claim it is one of the most extreme molecular complexes of the Milky Way.
\item Located at only 6~kpc from the Sun, W43 is an excellent laboratory to study the star formation process. The \emph{Herschel} Key Program Hi-GAL has already observed this region (e.g. Molinari et al. 2010) and a detailed analysis of the star formation content of this molecular cloud is ongoing.
\item With its location, extreme mass/density characteristics, and ab-normal velocity dispersion, W43 is the perfect laboratory to investigate the formation of such an extreme complex in the framework of the scenario of converging flows. The companion paper Motte et al. (in prep.) provides first signatures of colliding flows and poses the basis of further numerical simulations dedicated to W43. A large program with the IRAM 30~m ascertaining the diagnostics/signatures of colliding flows from large and low-density H~{\scriptsize I}/CO streams to small and high-density star formation seeds is ongoing (Motte, Schilke et al.). 
\end{enumerate}

\begin{acknowledgements}
   We thank Nemesio Rodriguez-Fernandez and Fran\c coise Combes for suggestions that improved the manuscript.  
  We thank Joe Hora for providing his \emph{Spitzer} images of Cygnus~X for comparative measurements, Thomas Dame for providing the $^{12}$CO~1--0  cubes of his Galactic plane survey and Marion Wienen for providing the NH$_{3}$~$\vlsr$ prior to publication. 
  Part of this work was supported by the ANR (\emph{Agence Nationale pour la Recherche}) project ``PROBeS", number ANR-08-BLAN-0241.
   LB acknowledges support from CONICYT projects  FONDAP 15010003 and Basal PFB-06.
\end{acknowledgements}


\bibliographystyle{aa}
\bibliography{w43reference}

\begin{thebibliography}{66}
\expandafter\ifx\csname natexlab\endcsname\relax\def\natexlab#1{#1}\fi

\bibitem[{{Anderson} \& {Bania}(2009)}]{anderson09}
{Anderson}, L.~D. \& {Bania}, T.~M. 2009, \apj, 690, 706

\bibitem[{{Bally} {et~al.}(2010){Bally}, {Anderson}, {Battersby}, {Calzoletti},
  {Digiorgio}, {Faustini}, {Ginsburg}, {Li}, {Nguyen-Luong}, {Molinari},
  {Motte}, {Pestalozzi}, {Plume}, {Rodon}, {Schilke}, {Schlingman},
  {Schneider-Bontemps}, {Shirley}, {Stringfellow}, {Testi}, {Traficante},
  {Veneziani}, \& {Zavagno}}]{bally10a}
{Bally}, J., {Anderson}, L.~D., {Battersby}, C., {et~al.} 2010, \aap, 518, L90+

\bibitem[{{Benjamin} {et~al.}(2003){Benjamin}, {Churchwell}, {Babler}, {Bania},
  {Clemens}, {Cohen}, {Dickey}, {Indebetouw}, {Jackson}, {Kobulnicky},
  {Lazarian}, {Marston}, {Mathis}, {Meade}, {Seager}, {Stolovy}, {Watson},
  {Whitney}, {Wolff}, \& {Wolfire}}]{benjamin03}
{Benjamin}, R.~A., {Churchwell}, E., {Babler}, B.~L., {et~al.} 2003, \pasp,
  115, 953

\bibitem[{{Benjamin} {et~al.}(2005){Benjamin}, {Churchwell}, {Babler},
  {Indebetouw}, {Meade}, {Whitney}, {Watson}, {Wolfire}, {Wolff}, {Ignace},
  {Bania}, {Bracker}, {Clemens}, {Chomiuk}, {Cohen}, {Dickey}, {Jackson},
  {Kobulnicky}, {Mercer}, {Mathis}, {Stolovy}, \&
  {Uzpen}}]{2005ApJ...630L.149B}
{Benjamin}, R.~A., {Churchwell}, E., {Babler}, B.~L., {et~al.} 2005, \apjl,
  630, L149

\bibitem[{{Beuther} {et~al.}(2007){Beuther}, {Zhang}, {Bergin}, {Sridharan},
  {Hunter}, \& {Leurini}}]{beuther07}
{Beuther}, H., {Zhang}, Q., {Bergin}, E.~A., {et~al.} 2007, \aap, 468, 1045

\bibitem[{{Bloemen} {et~al.}(1986){Bloemen}, {Strong}, {Mayer-Hasselwander},
  {Blitz}, {Cohen}, {Dame}, {Grabelsky}, {Thaddeus}, {Hermsen}, \&
  {Lebrun}}]{bloemen86}
{Bloemen}, J.~B.~G.~M., {Strong}, A.~W., {Mayer-Hasselwander}, H.~A., {et~al.}
  1986, \aap, 154, 25

\bibitem[{{Blum} {et~al.}(1999){Blum}, {Damineli}, \& {Conti}}]{blum99}
{Blum}, R.~D., {Damineli}, A., \& {Conti}, P.~S. 1999, \aj, 117, 1392

\bibitem[{{Brand} \& {Blitz}(1993)}]{brand93}
{Brand}, J. \& {Blitz}, L. 1993, \aap, 275, 67

\bibitem[{{Burton}(1970)}]{1970A&AS....2..291B}
{Burton}, W.~B. 1970, \aaps, 2, 291

\bibitem[{{Carpenter} \& {Sanders}(1998)}]{carpenter98}
{Carpenter}, J.~M. \& {Sanders}, D.~B. 1998, \aj, 116, 1856

\bibitem[{{Cesaroni} {et~al.}(1994){Cesaroni}, {Churchwell}, {Hofner},
  {Walmsley}, \& {Kurtz}}]{cesaroni94}
{Cesaroni}, R., {Churchwell}, E., {Hofner}, P., {Walmsley}, C.~M., \& {Kurtz},
  S. 1994, \aap, 288, 903

\bibitem[{{Cesaroni} {et~al.}(1998){Cesaroni}, {Hofner}, {Walmsley}, \&
  {Churchwell}}]{cesaroni98}
{Cesaroni}, R., {Hofner}, P., {Walmsley}, C.~M., \& {Churchwell}, E. 1998,
  \aap, 331, 709

\bibitem[{{Churchwell} {et~al.}(2009){Churchwell}, {Babler}, {Meade},
  {Whitney}, {Benjamin}, {Indebetouw}, {Cyganowski}, {Robitaille}, {Povich},
  {Watson}, \& {Bracker}}]{churchwell09}
{Churchwell}, E., {Babler}, B.~L., {Meade}, M.~R., {et~al.} 2009, \pasp, 121,
  213

\bibitem[{{Comer\'on} \& {Torra}(1994)}]{comeron94}
{Comer\'on}, F. \& {Torra}, J. 1994, \aap, 281, 35

\bibitem[{{Dahmen} {et~al.}(1998){Dahmen}, {Huttemeister}, {Wilson}, \&
  {Mauersberger}}]{1998A&A...331..959D}
{Dahmen}, G., {Huttemeister}, S., {Wilson}, T.~L., \& {Mauersberger}, R. 1998,
  \aap, 331, 959

\bibitem[{{Dame} {et~al.}(1986){Dame}, {Elmegreen}, {Cohen}, \&
  {Thaddeus}}]{dame86}
{Dame}, T.~M., {Elmegreen}, B.~G., {Cohen}, R.~S., \& {Thaddeus}, P. 1986,
  \apj, 305, 892

\bibitem[{{Dame} {et~al.}(2001){Dame}, {Hartmann}, \& {Thaddeus}}]{dame01}
{Dame}, T.~M., {Hartmann}, D., \& {Thaddeus}, P. 2001, \apj, 547, 792

\bibitem[{{Elmegreen} \& {Elmegreen}(1987)}]{elmegreen87}
{Elmegreen}, B.~G. \& {Elmegreen}, D.~M. 1987, \apj, 320, 182

\bibitem[{{Fukui} {et~al.}(2009){Fukui}, {Kawamura}, {Wong}, {Murai},
  {Iritani}, {Mizuno}, {Mizuno}, {Onishi}, {Hughes}, {Ott}, {Muller},
  {Staveley-Smith}, \& {Kim}}]{fukui09}
{Fukui}, Y., {Kawamura}, A., {Wong}, T., {et~al.} 2009, \apj, 705, 144

\bibitem[{{Graf} {et~al.}(1998){Graf}, {Haas}, {Honingh}, {Jacobs}, {Schieder},
  \& {Stutzki}}]{graf98}
{Graf}, U.~U., {Haas}, S., {Honingh}, C.~E., {et~al.} 1998, in Society of
  Photo-Optical Instrumentation Engineers (SPIE) Conference Series, Vol. 3357,
  Society of Photo-Optical Instrumentation Engineers (SPIE) Conference Series,
  ed. {T.~G.~Phillips}, 159--166

\bibitem[{{Hammersley} {et~al.}(2000){Hammersley}, {Garz{\'o}n}, {Mahoney},
  {L{\'o}pez-Corredoira}, \& {Torres}}]{HH00}
{Hammersley}, P.~L., {Garz{\'o}n}, F., {Mahoney}, T.~J.,
  {L{\'o}pez-Corredoira}, M., \& {Torres}, M.~A.~P. 2000, \mnras, 317, L45

\bibitem[{{Heitsch} \& {Hartmann}(2008)}]{heitsch08}
{Heitsch}, F. \& {Hartmann}, L. 2008, \apj, 689, 290

\bibitem[{{Hennebelle} {et~al.}(2008){Hennebelle}, {Banerjee},
  {V{\'a}zquez-Semadeni}, {Klessen}, \& {Audit}}]{hennebelle08}
{Hennebelle}, P., {Banerjee}, R., {V{\'a}zquez-Semadeni}, E., {Klessen}, R.~S.,
  \& {Audit}, E. 2008, \aap, 486, L43

\bibitem[{{Heyer} {et~al.}(2001){Heyer}, {Carpenter}, \& {Snell}}]{heyer01}
{Heyer}, M.~H., {Carpenter}, J.~M., \& {Snell}, R.~L. 2001, \apj, 551, 852

\bibitem[{{Hora} {et~al.}(2009){Hora}, {Bontemps}, {Megeath}, {Schneider},
  {Motte}, {Carey}, {Simon}, {Keto}, {Smith}, {Allen}, {Gutermuth}, {Fazio},
  {Adams}, \& {Cygnus-X 24 Micron Data Processing Team}}]{hora09}
{Hora}, J.~L., {Bontemps}, S., {Megeath}, S.~T., {et~al.} 2009, in Bulletin of
  the American Astronomical Society, Vol.~41, Bulletin of the American
  Astronomical Society, 498--498

\bibitem[{{Jackson} {et~al.}(2006){Jackson}, {Rathborne}, {Shah}, {Simon},
  {Bania}, {Clemens}, {Chambers}, {Johnson}, {Dormody}, {Lavoie}, \&
  {Heyer}}]{jackson06}
{Jackson}, J.~M., {Rathborne}, J.~M., {Shah}, R.~Y., {et~al.} 2006, \apjs, 163,
  145

\bibitem[{{Kennicutt}(1998)}]{kennicutt98}
{Kennicutt}, Jr., R.~C. 1998, \apj, 498, 541

\bibitem[{{Koda} {et~al.}(2009){Koda}, {Scoville}, {Sawada}, {La Vigne},
  {Vogel}, {Potts}, {Carpenter}, {Corder}, {Wright}, {White}, {Zauderer},
  {Patience}, {Sargent}, {Bock}, {Hawkins}, {Hodges}, {Kemball}, {Lamb},
  {Plambeck}, {Pound}, {Scott}, {Teuben}, \& {Woody}}]{koda09}
{Koda}, J., {Scoville}, N., {Sawada}, T., {et~al.} 2009, \apjl, 700, L132

\bibitem[{{Kramer} {et~al.}(1998){Kramer}, {Degiacomi}, {Graf}, {Hills},
  {Miller}, {Schieder}, {Schneider}, {Stutzki}, \& {Winnewisser}}]{kramer98b}
{Kramer}, C., {Degiacomi}, C.~G., {Graf}, U.~U., {et~al.} 1998, in Society of
  Photo-Optical Instrumentation Engineers (SPIE) Conference Series, Vol. 3357,
  Society of Photo-Optical Instrumentation Engineers (SPIE) Conference Series,
  ed. {T.~G.~Phillips}, 711--720

\bibitem[{{Kroupa}(2001)}]{kroupa01}
{Kroupa}, P. 2001, \mnras, 322, 231

\bibitem[{{Kuno} {et~al.}(1995){Kuno}, {Nakai}, {Handa}, \& {Sofue}}]{kuno95}
{Kuno}, N., {Nakai}, N., {Handa}, T., \& {Sofue}, Y. 1995, \pasj, 47, 745

\bibitem[{{Lester} {et~al.}(1985){Lester}, {Dinerstein}, {Werner}, {Harvey},
  {Evans}, \& {Brown}}]{Lester85}
{Lester}, D.~F., {Dinerstein}, H.~L., {Werner}, M.~W., {et~al.} 1985, \apj,
  296, 565

\bibitem[{{Liszt}(1995)}]{liszt95}
{Liszt}, H.~S. 1995, \aj, 109, 1204

\bibitem[{{L{\'o}pez-Corredoira} {et~al.}(2007){L{\'o}pez-Corredoira},
  {Cabrera-Lavers}, {Mahoney}, {Hammersley}, {Garz{\'o}n}, \&
  {Gonz{\'a}lez-Fern{\'a}ndez}}]{2007AJ....133..154L}
{L{\'o}pez-Corredoira}, M., {Cabrera-Lavers}, A., {Mahoney}, T.~J., {et~al.}
  2007, \aj, 133, 154

\bibitem[{{Martin} \& {Friedli}(1997)}]{1997A&A...326..449M}
{Martin}, P. \& {Friedli}, D. 1997, \aap, 326, 449

\bibitem[{{Molinari} {et~al.}(2010){Molinari}, {Swinyard}, {Bally}, {Barlow},
  {Bernard}, {Martin}, {Moore}, {Noriega-Crespo}, {Plume}, {Testi}, {Zavagno},
  {Abergel}, {Ali}, {Anderson}, {Andr{\'e}}, {Baluteau}, {Battersby},
  {Beltr{\'a}n}, {Benedettini}, {Billot}, {Blommaert}, {Bontemps}, {Boulanger},
  {Brand}, {Brunt}, {Burton}, {Calzoletti}, {Carey}, {Caselli}, {Cesaroni},
  {Cernicharo}, {Chakrabarti}, {Chrysostomou}, {Cohen}, {Compiegne}, {de
  Bernardis}, {de Gasperis}, {di Giorgio}, {Elia}, {Faustini}, {Flagey},
  {Fukui}, {Fuller}, {Ganga}, {Garcia-Lario}, {Glenn}, {Goldsmith}, {Griffin},
  {Hoare}, {Huang}, {Ikhenaode}, {Joblin}, {Joncas}, {Juvela}, {Kirk},
  {Lagache}, {Li}, {Lim}, {Lord}, {Marengo}, {Marshall}, {Masi}, {Massi},
  {Matsuura}, {Minier}, {Miville-Desch{\^e}nes}, {Montier}, {Morgan}, {Motte},
  {Mottram}, {M{\"u}ller}, {Natoli}, {Neves}, {Olmi}, {Paladini}, {Paradis},
  {Parsons}, {Peretto}, {Pestalozzi}, {Pezzuto}, {Piacentini}, {Piazzo},
  {Polychroni}, {Pomar{\`e}s}, {Popescu}, {Reach}, {Ristorcelli}, {Robitaille},
  {Robitaille}, {Rod{\'o}n}, {Roy}, {Royer}, {Russeil}, {Saraceno}, {Sauvage},
  {Schilke}, {Schisano}, {Schneider}, {Schuller}, {Schulz}, {Sibthorpe},
  {Smith}, {Smith}, {Spinoglio}, {Stamatellos}, {Strafella}, {Stringfellow},
  {Sturm}, {Taylor}, {Thompson}, {Traficante}, {Tuffs}, {Umana}, {Valenziano},
  {Vavrek}, {Veneziani}, {Viti}, {Waelkens}, {Ward-Thompson}, {White},
  {Wilcock}, {Wyrowski}, {Yorke}, \& {Zhang}}]{molinari10}
{Molinari}, S., {Swinyard}, B., {Bally}, J., {et~al.} 2010, \aap, 518, L100+

\bibitem[{{Mooney} {et~al.}(1995){Mooney}, {Sievers}, {Mezger}, {Solomon},
  {Kreysa}, {Haslam}, \& {Lemke}}]{mooney95}
{Mooney}, T., {Sievers}, A., {Mezger}, P.~G., {et~al.} 1995, \aap, 299, 869

\bibitem[{{Motte} {et~al.}(2007){Motte}, {Bontemps}, {Schilke}, {Schneider},
  {Menten}, \& {Brogui{\`e}re}}]{motte07}
{Motte}, F., {Bontemps}, S., {Schilke}, P., {et~al.} 2007, \aap, 476, 1243

\bibitem[{{Motte} {et~al.}(2003){Motte}, {Schilke}, \& {Lis}}]{motte03}
{Motte}, F., {Schilke}, P., \& {Lis}, D.~C. 2003, \apj, 582, 277

\bibitem[{{Ossenkopf} \& {Henning}(1994)}]{Ossenkopf94}
{Ossenkopf}, V. \& {Henning}, T. 1994, \aap, 291, 943

\bibitem[{{Pandian} {et~al.}(2008){Pandian}, {Momjian}, \&
  {Goldsmith}}]{pandian08}
{Pandian}, J.~D., {Momjian}, E., \& {Goldsmith}, P.~F. 2008, \aap, 486, 191

\bibitem[{{Peeters} {et~al.}(2004){Peeters}, {Spoon}, \& {Tielens}}]{peeters04}
{Peeters}, E., {Spoon}, H.~W.~W., \& {Tielens}, A.~G.~G.~M. 2004, \apj, 613,
  986

\bibitem[{{Pound} \& {Blitz}(1993)}]{pound93}
{Pound}, M.~W. \& {Blitz}, L. 1993, \apj, 418, 328

\bibitem[{{Pratap} {et~al.}(1999){Pratap}, {Megeath}, \& {Bergin}}]{pratap99}
{Pratap}, P., {Megeath}, S.~T., \& {Bergin}, E.~A. 1999, \apj, 517, 799

\bibitem[{{Rand}(1993)}]{rand93b}
{Rand}, R.~J. 1993, \apj, 410, 68

\bibitem[{{Rand} \& {Kulkarni}(1990)}]{rand90}
{Rand}, R.~J. \& {Kulkarni}, S.~R. 1990, \apjl, 349, L43

\bibitem[{{Rathborne} {et~al.}(2009){Rathborne}, {Johnson}, {Jackson}, {Shah},
  \& {Simon}}]{rathborne09}
{Rathborne}, J.~M., {Johnson}, A.~M., {Jackson}, J.~M., {Shah}, R.~Y., \&
  {Simon}, R. 2009, \apjs, 182, 131

\bibitem[{{Reid} {et~al.}(2009){Reid}, {Menten}, {Zheng}, {Brunthaler},
  {Moscadelli}, {Xu}, {Zhang}, {Sato}, {Honma}, {Hirota}, {Hachisuka}, {Choi},
  {Moellenbrock}, \& {Bartkiewicz}}]{2009ApJ...700..137R}
{Reid}, M.~J., {Menten}, K.~M., {Zheng}, X.~W., {et~al.} 2009, \apj, 700, 137

\bibitem[{{Rodriguez-Fernandez} \& {Combes}(2008)}]{rodriguez08}
{Rodriguez-Fernandez}, N.~J. \& {Combes}, F. 2008, \aap, 489, 115

\bibitem[{{Rohlfs} \& {Wilson}(2000)}]{rohlfs00}
{Rohlfs}, K. \& {Wilson}, T.~L. 2000, {Tools of radio astronomy} (Springer)

\bibitem[{{Roman-Duval} {et~al.}(2009){Roman-Duval}, {Jackson}, {Heyer},
  {Johnson}, {Rathborne}, {Shah}, \& {Simon}}]{roman09}
{Roman-Duval}, J., {Jackson}, J.~M., {Heyer}, M., {et~al.} 2009, \apj, 699,
  1153

\bibitem[{{Russeil}(2003)}]{russeil03}
{Russeil}, D. 2003, \aap, 397, 133

\bibitem[{{Russeil} {et~al.}(2011){Russeil}, {Pestalozzi}, {Mottram},
  {Bontemps}, {Anderson}, {Zavagno}, {Beltr{\'a}n}, {Bally}, {Brand}, {Brunt},
  {Cesaroni}, {Joncas}, {Marshall}, {Martin}, {Massi}, {Molinari}, {Moore},
  {Noriega-Crespo}, {Olmi}, {Thompson}, {Wienen}, \&
  {Wyrowski}}]{2011A&A...526A.151R}
{Russeil}, D., {Pestalozzi}, M., {Mottram}, J.~C., {et~al.} 2011, \aap, 526,
  A151+

\bibitem[{{Schneider} {et~al.}(2006){Schneider}, {Bontemps}, {Simon}, {Jakob},
  {Motte}, {Miller}, {Kramer}, \& {Stutzki}}]{schneider06}
{Schneider}, N., {Bontemps}, S., {Simon}, R., {et~al.} 2006, \aap, 458, 855

\bibitem[{{Schuller} {et~al.}(2009){Schuller}, {Menten}, {Contreras},
  {Wyrowski}, {Schilke}, {Bronfman}, {Henning}, {Walmsley}, {Beuther},
  {Bontemps}, {Cesaroni}, {Deharveng}, {Garay}, {Herpin}, {Lefloch}, {Linz},
  {Mardones}, {Minier}, {Molinari}, {Motte}, {Nyman}, {Reveret}, {Risacher},
  {Russeil}, {Schneider}, {Testi}, {Troost}, {Vasyunina}, {Wienen}, {Zavagno},
  {Kovacs}, {Kreysa}, {Siringo}, \& {Wei{\ss}}}]{schuller09}
{Schuller}, F., {Menten}, K.~M., {Contreras}, Y., {et~al.} 2009, \aap, 504, 415

\bibitem[{{Silk}(1997)}]{silk97}
{Silk}, J. 1997, \apj, 481, 703

\bibitem[{{Simon} {et~al.}(2001){Simon}, {Jackson}, {Clemens}, {Bania}, \&
  {Heyer}}]{simon01}
{Simon}, R., {Jackson}, J.~M., {Clemens}, D.~P., {Bania}, T.~M., \& {Heyer},
  M.~H. 2001, \apj, 551, 747

\bibitem[{{Smith} {et~al.}(1978){Smith}, {Biermann}, \& {Mezger}}]{smith78}
{Smith}, L.~F., {Biermann}, P., \& {Mezger}, P.~G. 1978, \aap, 66, 65

\bibitem[{{Solomon} {et~al.}(1987){Solomon}, {Rivolo}, {Barrett}, \&
  {Yahil}}]{solomon87}
{Solomon}, P.~M., {Rivolo}, A.~R., {Barrett}, J., \& {Yahil}, A. 1987, \apj,
  319, 730

\bibitem[{{Spitzer}(1978)}]{spitzer78}
{Spitzer}, L. 1978, {Physical processes in the interstellar medium} (Wiley VCH)

\bibitem[{{Starck} \& {Murtagh}(2006)}]{starck06}
{Starck}, J. \& {Murtagh}, F. 2006, {Astronomical Image and Data Analysis}
  (Springer)

\bibitem[{{Stil} {et~al.}(2006){Stil}, {Taylor}, {Dickey}, {Kavars}, {Martin},
  {Rothwell}, {Boothroyd}, {Lockman}, \& {McClure-Griffiths}}]{stil06}
{Stil}, J.~M., {Taylor}, A.~R., {Dickey}, J.~M., {et~al.} 2006, \aj, 132, 1158

\bibitem[{{V{\'a}zquez-Semadeni} {et~al.}(2007){V{\'a}zquez-Semadeni},
  {G{\'o}mez}, {Jappsen}, {Ballesteros-Paredes}, {Gonz{\'a}lez}, \&
  {Klessen}}]{vazquez07}
{V{\'a}zquez-Semadeni}, E., {G{\'o}mez}, G.~C., {Jappsen}, A.~K., {et~al.}
  2007, \apj, 657, 870

\bibitem[{{Watson} \& {Hanson}(1997)}]{watson97}
{Watson}, A.~M. \& {Hanson}, M.~M. 1997, \apjl, 490, L165+

\bibitem[{{Wood} \& {Churchwell}(1989)}]{wood89}
{Wood}, D.~O.~S. \& {Churchwell}, E. 1989, \apj, 340, 265

\bibitem[{{Wu} {et~al.}(2005){Wu}, {Cao}, {Hao}, {Liu}, {Wang}, {Xia}, {Deng},
  \& {Young}}]{wu05}
{Wu}, H., {Cao}, C., {Hao}, C., {et~al.} 2005, \apjl, 632, L79

\end{thebibliography}

\begin{appendix}
\section{KOSMA $^{12}$CO~2--1 and $^{12}$CO~3--2 imaging of the W43 molecular cloud complex} 

\begin{figure*}[h]
 \vskip -0.5cm
  \centering
  \includegraphics[angle=-90,width=15cm]{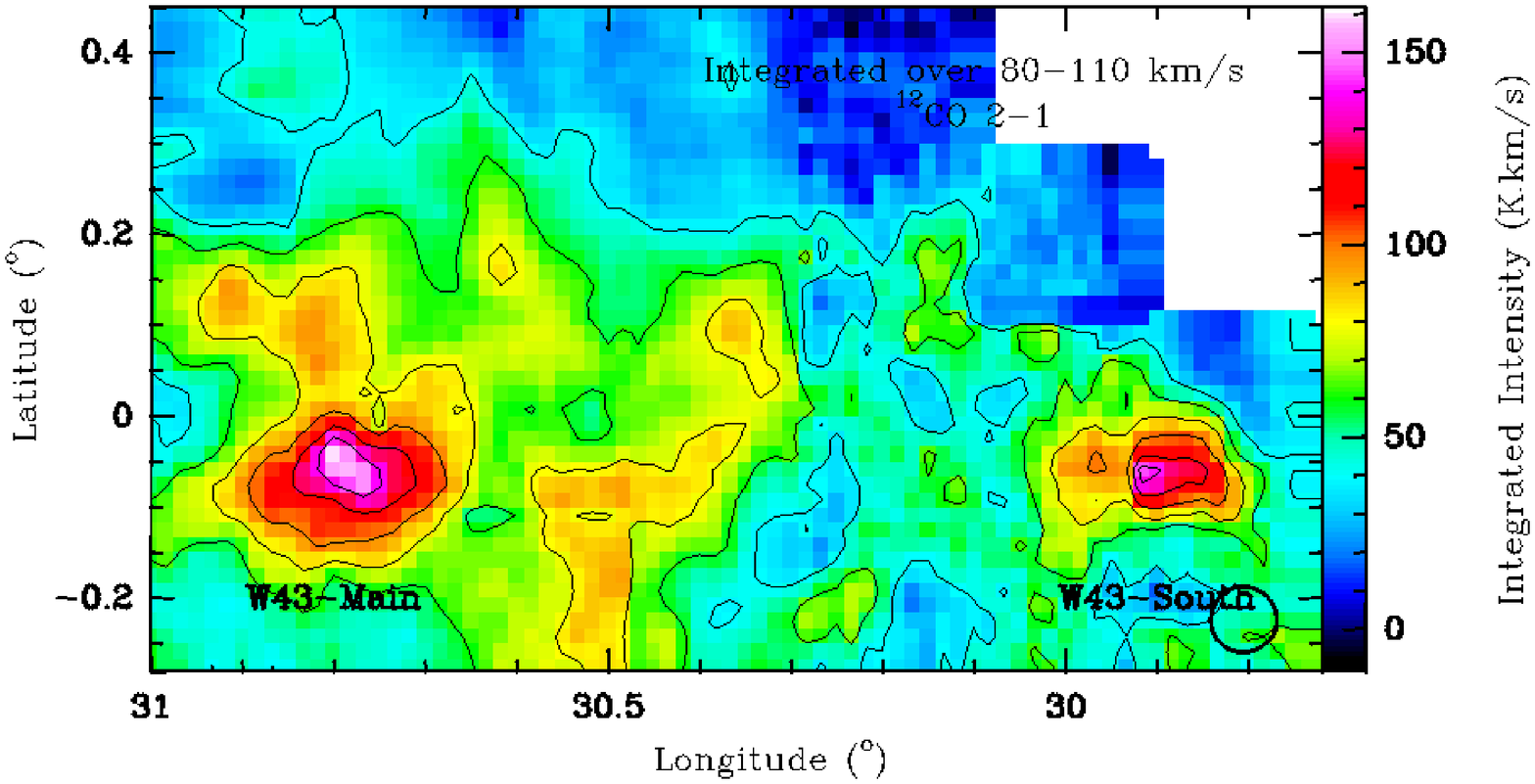}   
\caption{$^{12}$CO~2--1 intensity map of the inner part of the W43 molecular complex obtained with KOSMA. The lines are integrated over the $80-110~\kms$ velocity range. Contours go from to 40 to 140 by 20~in unit of K$~\kms$. The HPBW is plotted in the bottom-left corner.} 
\label{12co21chanful}
 \label{12co21int}
   \end{figure*}

\begin{figure*}[h]
 \vskip -0.5cm
  \centering
     \includegraphics[angle=-90,width=15cm]{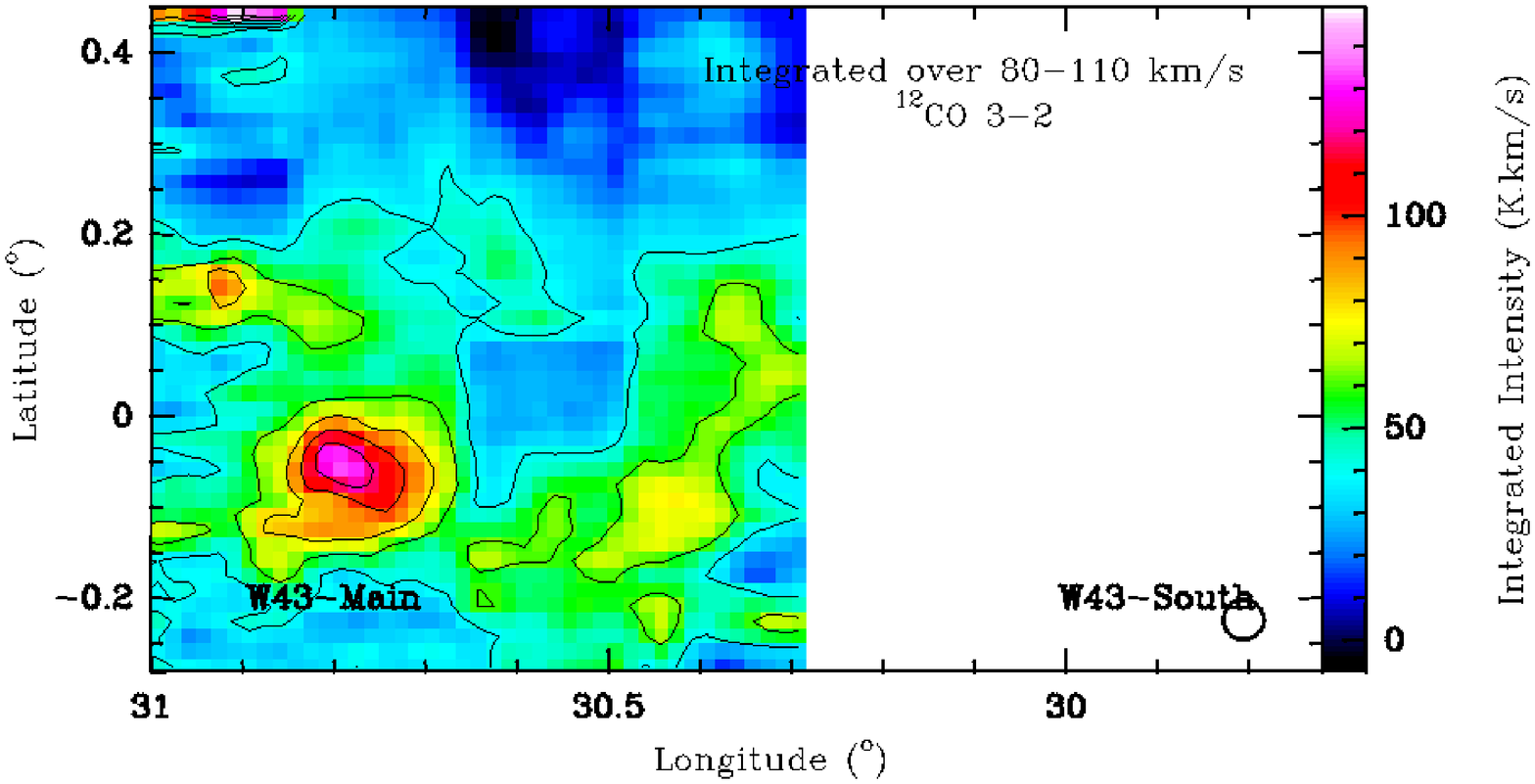}
\caption{$^{12}$CO~3--2 intensity map of the inner part of the W43 molecular complex obtained with KOSMA. The lines are integrated over the $80-110~\kms$ velocity range. Contours go from to 40 to 140 by 20~in unit of K$~\kms$. The HPBW is plotted in the bottom-left corner.} 
   \end{figure*} 
   
\begin{figure*}[t]
 \vskip -2.3cm
  \centering
  \includegraphics[angle=0,height=25.5cm]{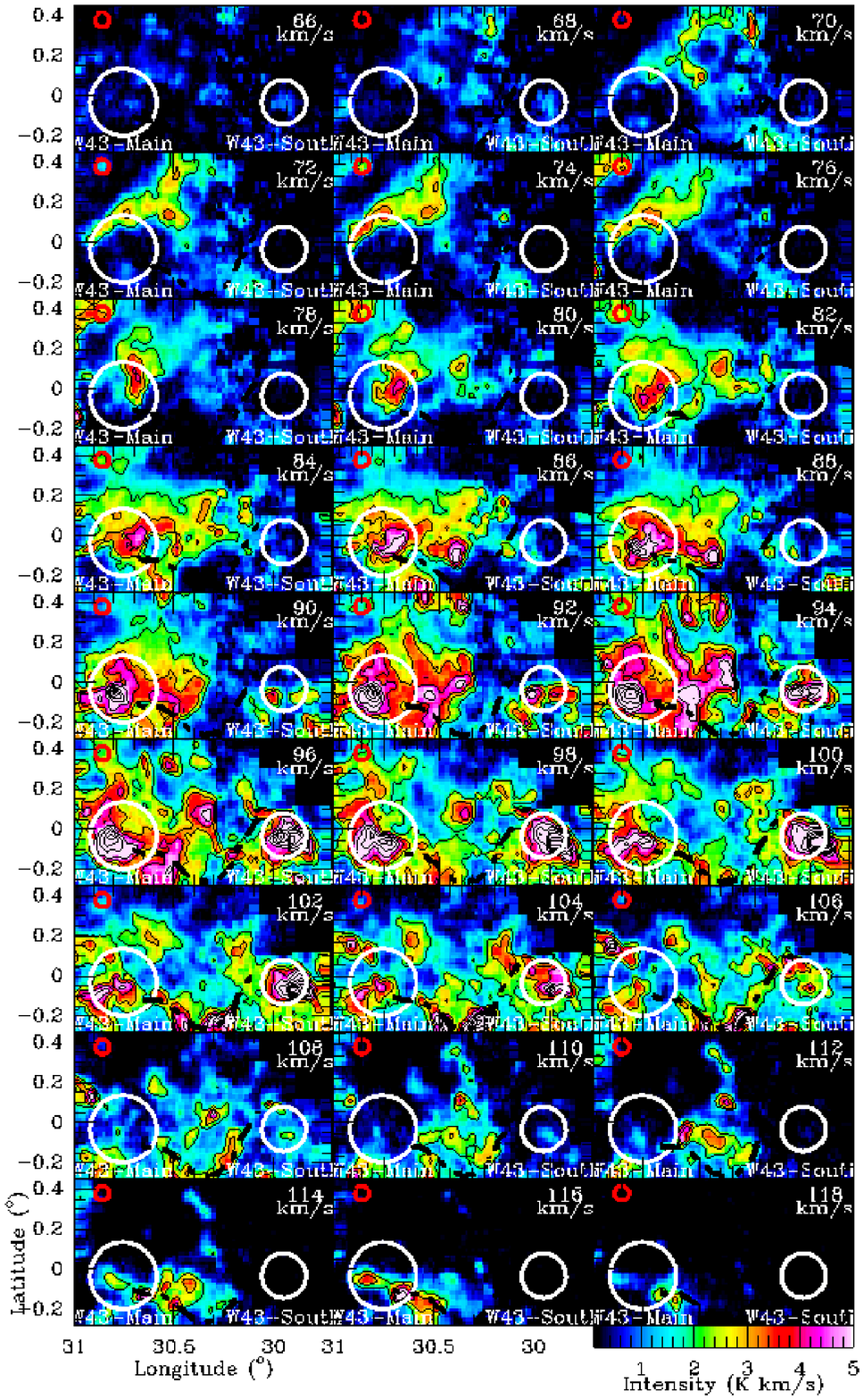} 
 \vskip -0.3cm
\caption{Velocity maps of the inner part of the W43 molecular complex in the $^{12}$CO~2--1 emission obtained with KOSMA, integrated over $\sim 2~\kms$-wide channels
ranging from $\vlsr=66$ to $118~\kms$. The  W43-Main and W43-South regions are indicated with white circles. The main molecular bridge linking them is outlined by a dashed curve. The HPBW is plotted in the top-left corner. }
\label{12co21chanful}
 \vskip -1cm
   \end{figure*}

\begin{figure*}[t]
 \vskip -2.3cm
  \centering
  \includegraphics[angle=0,height=25.5cm]{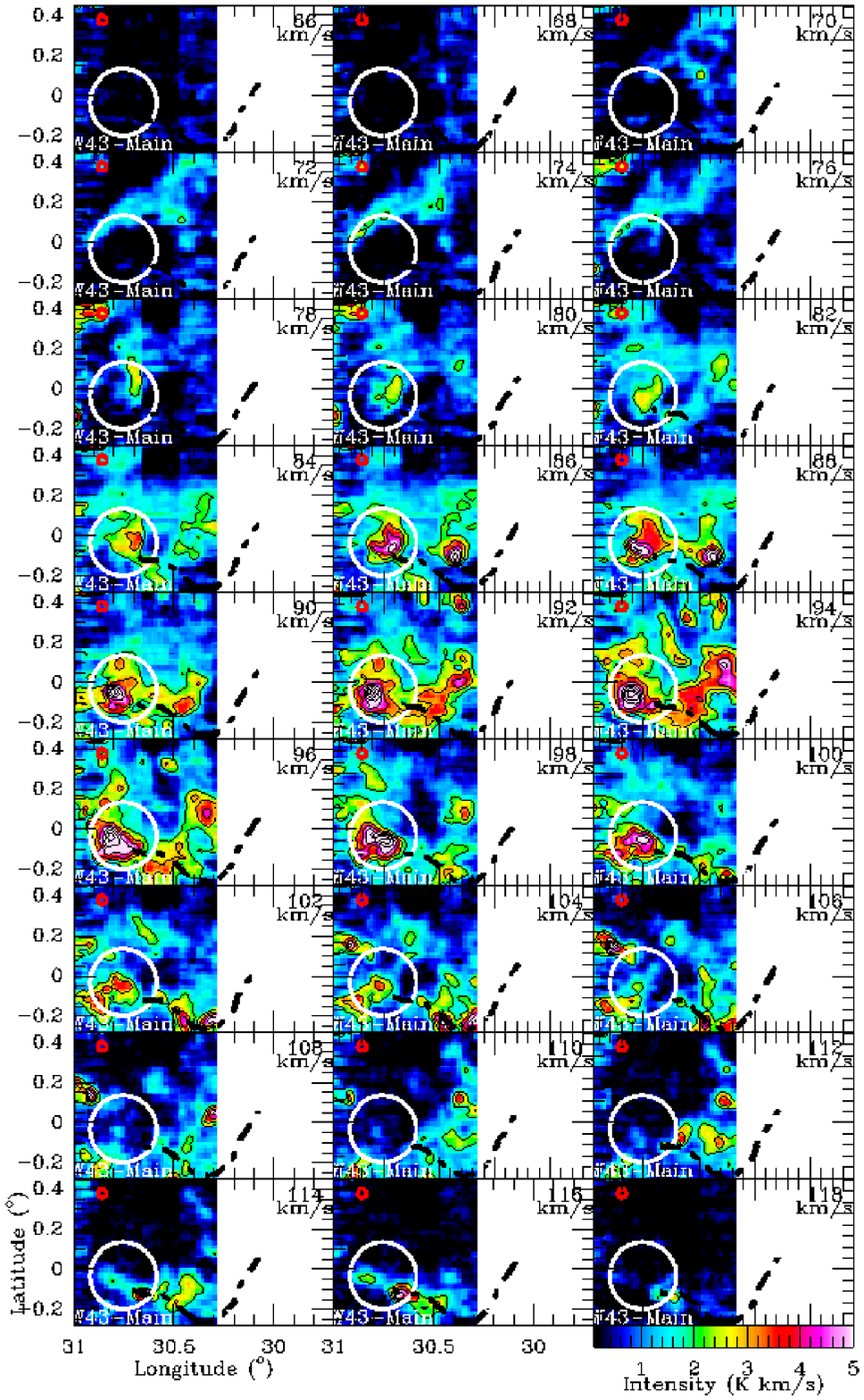} 
   \vskip -0.3cm
\caption{Velocity maps of the inner part of the W43 molecular complex in the $^{12}$CO~3--2  emission obtained with KOSMA, integrated over $\sim 2~\kms$-wide channels
ranging from $\vlsr=66$ to $118~\kms$. The  W43-Main and W43-South regions are indicated with white circles. The main molecular bridge linking them is outlined by a dashed curve. The HPBW is plotted in the top-left corner. }
\label{12co32chanful}
 \vskip -1cm
  \end{figure*}
\end{appendix}

\end{document}